\documentclass[12pt]{article}

\usepackage{newtxtext,newtxmath}

\usepackage{graphicx}
\usepackage{lineno}

\usepackage[a4paper, margin=2.5cm]{geometry}

\renewenvironment{abstract}
	{\quotation}
	{\endquotation}

\date{}

\renewcommand\refname{References}

\makeatletter
\renewcommand{\fnum@figure}{\textbf{Figure \thefigure}}
\renewcommand{\fnum@table}{\textbf{Table \thetable}}
\makeatother

\usepackage[super,comma,sort&compress]{natbib}

\usepackage{url}

\newcommand{\EuPdSi}{EuPd$_2$(Si$_{1-x}$Ge$_x$)$_2$}
\newcommand{\EuPdSiGex}{EuPd$_2$(Si$_{0.89}$Ge$_{0.11}$)$_2$}

\newcommand{\onlinecite}[1]{{\setcitestyle{numbers}[\cite{#1}]}}

\usepackage{hyperref}
\hypersetup{hidelinks} 

\def\scititle{
Elastic tricriticality at the collapse of magnetic order in a valence fluctuating Eu-based compound}
\title{\bfseries \boldmath \scititle}

\author{
	Yassine Agarmani$^{1\ast}$,
	Arif Ata$^{1}$,
    Lars Franke$^{2}$,
	Jan Zimmermann$^{1}$,
    Ulrich Tutsch$^{1}$,\and
    Marius Peters$^{1}$,
    Michelle Ocker$^{1}$,
    Kristin Kliemt$^{1}$,\and
    Cornelius Krellner$^{1}$,
    Bernd Wolf$^{1}$,
    Markus Garst$^{2}$,
    Michael Lang$^{1\ast}$\and
	\small$^{1}$Institute of Physics, Goethe University Frankfurt, 60438 Frankfurt (M), Germany\and
	\small$^{2}$Institute of Theoretical Solid State Physics, Karlsruhe Institute of Technology, 76131 Karlsruhe, Germany.\and
	\small$^\ast$Corresponding author. Email: agarmani@physik.uni-frankfurt.de \and
}

\begin{document} 

\maketitle
\begin{abstract} \bfseries \boldmath
Spin and charge fluctuations are key for understanding the properties of materials with strongly correlated electrons, but their coupling to the atomic crystal lattice is often assumed to be weak and perturbative. In this work, we demonstrate that at tricritical points (TCPs), that are associated with electronic correlations, a coupling to crystal elasticity is generically non-perturbative. It will modify the tricritical universality class leading to \textit{elastic tricriticality}, characterized by a compressibility that diverges in a universal manner with the correlation length. To test this scenario experimentally, we use single crystals of \EuPdSi\, where, depending on $x$ and pressure, magnetic order collapses due to strong valence fluctuations. First, we confirm the presence of a TCP for $x \approx 0.1$ by using complementary probes, in particular, measurements of relative length changes under control of temperature and He-gas pressure. Second, by studying the elastic response, we uncover clear signatures of elastic tricriticality, notably a drastically enhanced compressibility.
\end{abstract}

\noindent \textbf{Keywords: \\ Strongly correlated systems $|$ Critical phenomena $|$ Elasticity $|$ Rare-earth intermetallics}

\newpage


Materials with strongly correlated electrons are of continuing high interest due to the wealth of intriguing phenomena and the challenges they pose for the understanding of various instabilities and the role played by different types of fluctuations. Electron-electron correlations and the associated phase transitions can often be controlled by pressure. Key examples include the Mott metal-insulator transition \cite{Mott1990} revealed, e.g., in transition-metal oxides \cite{McWhan1969,Jayaraman1970} and organic charge-transfer salts \cite{Kanoda1997}, as well as the valence transition in certain rare earth- and actinide materials including the  $\gamma$ - $\alpha$ transition in cerium \cite{Jayaraman1965,Lawrence1975,Koskenmaki1978}, see, e.g.,  Refs.~\cite{Robinson1979,Parks1985} for early reviews. While these phenomena appear to be driven by strong electron-electron interactions, the high sensitivity to pressure indicates a pronounced coupling between electronic and lattice degrees of freedom.

\begin{figure}[t!]
\centering
	\includegraphics[width=0.75\textwidth]{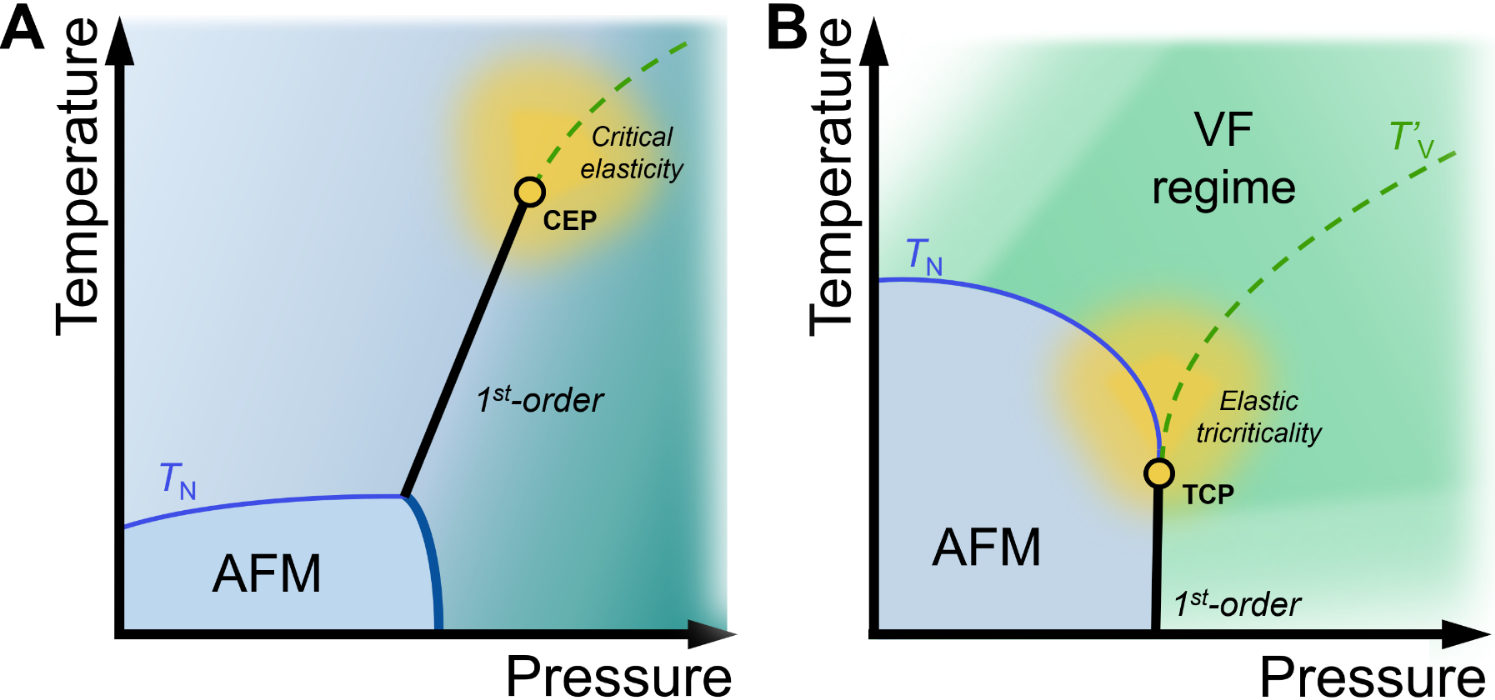}
	\caption{\textbf{Schematic temperature-pressure phase diagrams for materials with strongly correlated electrons featuring critical points.} (\textbf{A}) A line of first-order transitions terminates in a critical endpoint (CEP) out of which a crossover line (broken line) evolves. Around the CEP \textit{critical elasticity} (yellow-filled area) is expected. At low temperatures and low pressures, one often finds long-range antiferromagnetic (AFM) order. (\textbf{B}) A tricritical point (TCP) is revealed when a line of first-order transitions turns into a line of second-order transitions. This is accompanied by the appearance of a crossover line. As argued below, around the TCP one expects effects of \textit{elastic tricriticality} (yellow area). A scenario, implying a TCP, has been proposed in which AFM order collapses due to the effect of strong valence fluctuations (VF): The second-order transition associated with AFM order below $T_\mathrm{N}$ (blue line) turns into a first-order transition (black solid line) upon merging with the valence-crossover line $T'_\mathrm{V}$ (green dashed line).}
	\label{fig:Fig1new}
\end{figure}

Materials with Mott- or valence transitions have in common that their phase diagram contains a line of first-order phase transitions that terminates in a critical endpoint (CEP), see Fig.~\ref{fig:Fig1new}\,A. Close to the endpoint, the critical behavior is characterized by an emergent Ising order parameter $\psi$ similar to the liquid-gas endpoint \cite{Castellani1979,Kotliar2000}.  
However, in a solid this order parameter will involuntarily couple to the atomic crystal lattice, and this elastic coupling changes the universality class of the CEP resulting in \textit{critical elasticity} \cite{Cowley1976}. Instead of a liquid-gas CEP, the long-range shear forces of the crystal give rise to an elastic solid-solid CEP that is governed by mean-field behavior without microscopic critical fluctuations and, therefore, it lacks a diverging correlation length. For the valence transition, this has been pointed out and experimentally confirmed early on for the $\gamma$\,-\,$\alpha$ transition in Ce$_{1-x}$Th$_x$ \cite{Lawrence1975}, see also Refs.\cite{Dzero2010, Hackl2008a}. For the Mott transition, experimental evidence was found in (V$_{1-x}$Cr$_x$)$_2$O$_3$ \cite{Jayaraman1970} as well as in the organic conductor $\kappa$-(BEDT-TTF)$_2$Cu[N(CN)$_2$]Cl \cite{Gati2016}, see also the discussion in Refs.\,\cite{Majumdar1994, Hassan2005, Zacharias2012, Thyzel2025}.

A more intricate situation arises when the electronic valence fluctuations (VF) are accompanied by critical spin fluctuations due to the vicinity of long-range magnetic order \cite{Watanabe2010, Watanabe2011}. 
The interplay of these different types of fluctuations is considered key to understanding a variety of intriguing phenomena in correlated  electron 
systems,\,including different types of unconventional superconductivity (see, e.g., Ref.\,\cite{Yuan2003}). In particular, it has been predicted that weak AFM order can be destabilized by strong valence fluctuations \cite{Watanabe2011}, resulting in a tricritical point (TCP) as schematically shown in Fig.~\ref{fig:Fig1new}\,B. This scenario is also corroborated by Landau theory, see Sec.~H in the Supporting Information (SI) for details. At such a TCP, the second-order phase transition turns into a first-order transition accompanied by a crossover line.

In the present work, we aim to elucidate the interplay between valence, spin, and elastic fluctuations, with particular emphasis on how the elastic coupling affects tricritical behavior materializing close to the TCP in Fig.~\ref{fig:Fig1new}\,B. We argue theoretically and we provide experimental evidence that the elastic coupling also changes the universality class of the TCP resulting in \textit{elastic tricriticality}. In contrast to the elastic solid-solid CEP, however, critical microscopic fluctuations of the coupled electronic-elastic system are maintained at the elastic TCP, implying a diverging correlation length and universal critical exponents.

For the realization of a phase diagram with a TCP, Eu-based intermetallics may provide a suitable basis given their tendency to form charge-fluctuating, intermediate-valence states in close proximity to magnetic order \cite{Segre1982, Bauminger1973, Onuki2020, Lai2022}. A particularly interesting case has been revealed by \EuPdSi. Instead of a first-order transition, generally observed at the valence instability \cite{Segre1982, Wada1999, Wada1999a, Mitsuda2012, Onuki2016,  Lai2022}, EuPd$_2$Si$_2$ shows a continuous valence change \cite{Sampathkumaran1981, Batlogg1982, Kemly1985, Fedchenko2024} indicative of a crossover \cite{Wolf2023, Peters2023}. With increasing Ge substitution $x$, which acts as negative (chemical) pressure, the crossover temperature $T'_\mathrm{V}$ becomes reduced for small substitution levels ($x < 0.1$) before the material develops antiferromagnetic order at $x > 0.1$ \cite{Wolf2023, Peters2023}. Interestingly, studies on a single crystal with $x \approx$ 0.1 performed at various pressures suggest that by applying a tiny amount of pressure on the order of 10 MPa, i.e., 0.1 per mille of the value of typical bulk moduli of metals, the antiferromagnetic order collapses via a first-order phase transition giving way to a state with strong valence fluctuations \cite{Wolf2023}. These findings hint at the existence of a TCP, which, if confirmed, would open up the exciting possibility of exploring the impact of critical charge and spin fluctuations on the elastic properties and thus verifying elastic tricriticality.

The objectives of the present work are twofold. First, we confirm the presence of a TCP for $x \approx 0.1$ by using complementary probes.
Second, we uncover clear signatures of elastic tricriticality by studying the elastic response of the material finely tuned to its TCP. 
We performed measurements of expansivity, resistivity, and magnetometry both as a function of temperature, $T$, at constant pressure, $p$, and vice versa, as a function of $p$ at constant $T$. Our experiments were made possible, in particular, by recent technical developments that enable high-resolution measurements of relative length changes to be performed under continuous variation of He-gas pressure \cite{Manna2012, Agarmani2022}.

\section*{Results}
All samples used in this work were cut off from a single, large \EuPdSiGex\, crystal to minimize effects related to sample-to-sample variations (see Table\,S1 and Fig.\,S2 in SI Appendix). In the following, we present results of 
expansivity, resistivity, and magnetometry.
Given that the target material, \EuPdSi, crystallizes in the tetragonal ThCr$_2$Si$_2$ (122) structure \cite{Sampathkumaran1981} (Fig.\,S1), and shows strongly anisotropic lattice effects at the valence change for $x = 0$, which are essentially governed by changes in the in-plane $a$ axis \cite{Jhans1987,Peters2023} (see also Fig. S4\,A), our study focuses on probing the $a$ axis expansivity.

\subsection*{Identification of the tricritical point}
\begin{figure*}[t!]
\centering
	\includegraphics[width=0.75\textwidth]{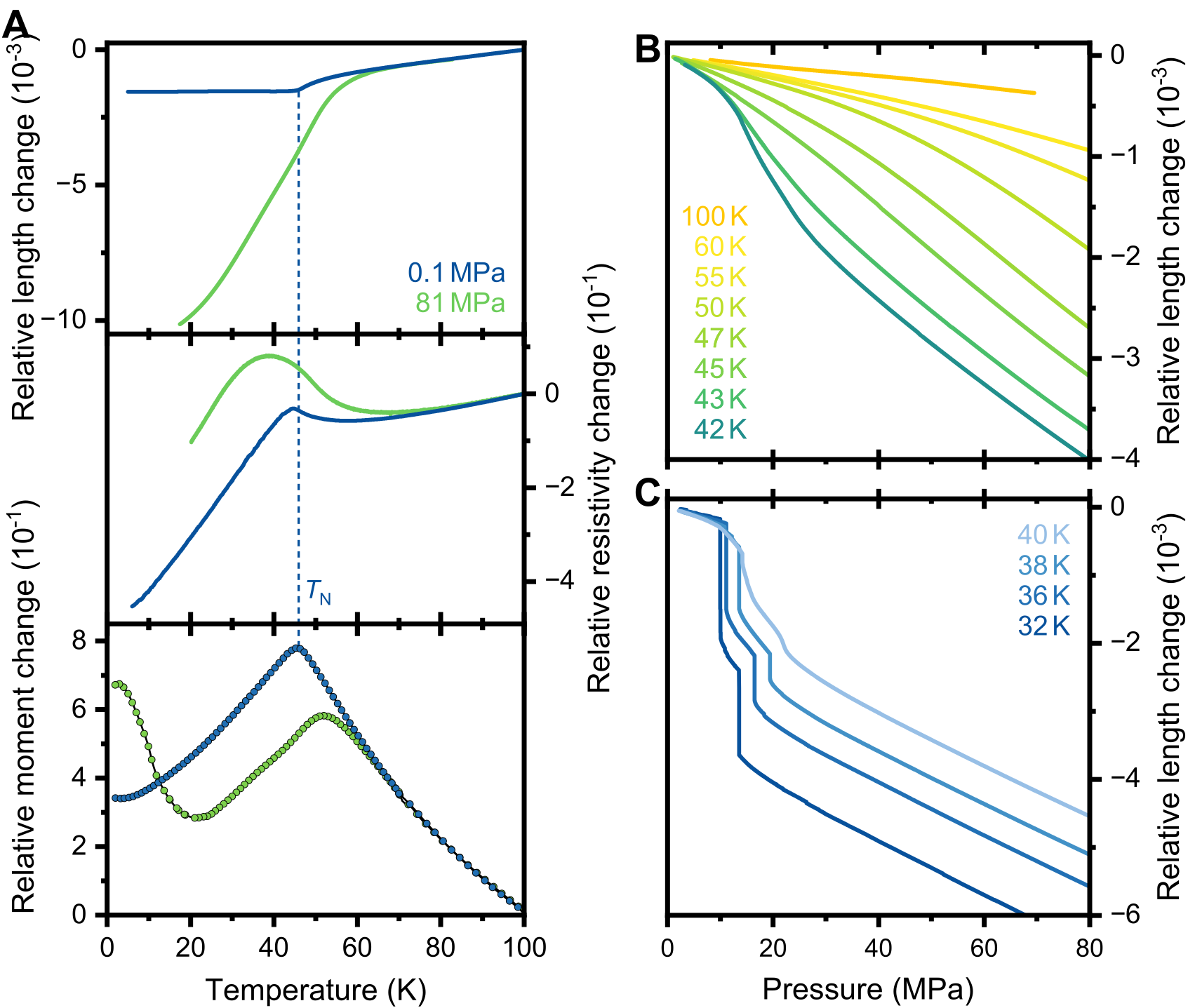}
	\caption{\textbf{Interplay between the antiferromagnetic order and valence fluctuations in EuPd$_{\mathbf{2}}$(Si$_{\mathbf{0.89}}$Ge$_{\mathbf{0.11}}$)$_\mathbf{2}$.} (\textbf{A}) Relative length change $\Delta L_a(T,p=\mathrm{const.})/L_a$ along the in-plane $a$ axis (top panel, sample \#2a), changes of the in-plane resistivity $\Delta \rho (T,p=\mathrm{const.})/\rho$ (middle panel, sample \#2a\_2) and changes of the magnetic moment $\Delta m_a(T,p=\mathrm{const.})/m_a$, with the magnetic field of $H=10$\,kOe\, oriented along the $a$ axis (bottom panel, composite sample consisting of \#2a\_2 - \#2a\_4), are shown as a function of temperature at two selected constant pressures of nominal 0.1 and 81\,MPa. The data were normalized to their values at 0.1\,MPa and 100\,K. Vertical broken line indicates the transition into AFM order at $T_\mathrm{N}$ under ambient-pressure ($p$ = 0.1\,MPa) conditions. (\textbf{B}) Relative length change $\Delta L_a(T=\mathrm{const.},p)/L_a$ as a function of pressure at selected temperatures above the tricritical point ($T>T_\mathrm{TCP}$), showing either a continuous phase transition, associated with antiferromagnetic order for $T<T_\mathrm{N}$, or a broad crossover behavior, associated with valence fluctuations for $T>T_\mathrm{N}$. (\textbf{C}) Relative length change $\Delta L_a(T=\mathrm{const.},p)/L_a$, at selected temperatures below the triciritcal point ($T<T_\mathrm{TCP}$), showing discontinuous length jumps reflecting a first-order phase transition. Two distinct jumps are observed, which are attributed to sample inhomogeneities, likely resulting from slightly different Ge concentrations with two accumulation points corresponding to two (majority and minority) volume fractions (see Sec. B in SI Appendix for details).}
	\label{fig:Fig2}
\end{figure*}

In Fig.\,\ref{fig:Fig2}\,A, we show exemplarily three data sets on \EuPdSiGex\, including the relative length change, $\Delta L_a(T,p=\mathrm{const.})/L_a$ (top panel), the change in resistivity, $\Delta  \rho(T,p=\mathrm{const.}) /\rho$ (middle panel), and the change in magnetic moment, $\Delta m_a(T,p=\mathrm{const.})/m_a$ (bottom panel), as a function of temperature, taken at two slightly different pressure values,
resulting in distinctly different ground states. At nominal ambient pressure (0.1\,MPa, blue curves), the material exhibits AFM order at low temperatures, as evidenced by the Curie-Weiss-like behavior in the magnetic susceptibility $\chi_a$ (= $m_a$/$H$, with $H$ the applied magnetic field) at elevated temperatures, followed by a sharp kink at $T_{\mathrm{N}} \approx 47$\,K (lower panel of Fig.\,\ref{fig:Fig2}\,A) below which a distinct magnetic (easy-plane) anisotropy evolves (see Fig.\,S5\,A), consistent with literature results \cite{Peters2023,Wolf2023}. Measurements of relative length changes (upper panel of Fig.\,\ref{fig:Fig2}\,A) reveal a small but accelerated decrease in $\Delta L_a(T,p=0.1\,\mathrm{MPa})/L_a$ upon cooling and a sharp kink at $47$\,K, reflecting a second-order phase transition at $T_{\mathrm{N}}$. For $T < T_{\mathrm{N}}$, $\Delta L_a(T,p=0.1\,\mathrm{MPa})/L_a$ becomes nearly temperature independent on the scale of Fig.\,\ref{fig:Fig2}\,A. In the resistivity at nominal 0.1\,MPa, we find a mild increase in approaching $T_{\mathrm{N}}$ from above, followed by a kink slightly below $T_{\mathrm{N}}$ and a rapid reduction for $T < T_{\mathrm{N}}$. 

A significantly different behavior is revealed by applying pressure as small as $p$ = 81\,MPa (green curves in Fig.\,\ref{fig:Fig2}\,A). The difference from the data at 0.1\,MPa begins to become visible around 70\,K, below which the sample length reveals a more rapid contraction. Upon further cooling, this is followed by a drastic, though continuous, change in $\Delta L_a(T,p=81\,\mathrm{MPa})/L_a$, totaling an overall contraction of about one percent. As demonstrated for non-substituted EuPd$_2$Si$_2$ (Fig.\,S4\,A), these are clear signatures of valence-crossover behavior. In the resistivity taken at nominal $p$ = 81\,MPa, the valence-crossover is reflected by a gradual increase followed by a broad maximum. The valence-crossover temperature $T'_\mathrm{V}$ can be identified with the inflection point on the high-temperature flank of this maximum (Fig.\,\ref{fig:Fig2}\,A, middle panel), which coincides with the point of maximum slope in  $\Delta L_a(T,p=81\,\mathrm{MPa})/L_a$ (see Figs.\,S6\,(B) and (C)). In the magnetic properties, the valence crossover manifests itself in deviations from Curie-Weiss-like behavior at elevated temperatures, followed by a rounded maximum and a low-temperature upturn (lower panel of Fig.\,\ref{fig:Fig2}\,A). These features, along with the lack of anisotropy, were identified as signatures of the VF regime characterized by a maximum in $d(\chi \cdot T)/dT$ at $T'_\mathrm{V}$ \cite{Wolf2023} (see also Fig.\,S6\,A).

The markedly different low-$T$ behavior revealed for \EuPdSiGex\, by slightly increasing the pressure from 0.1 to 81\,MPa (Fig.\,\ref{fig:Fig2}\,A), together with estimates of the electronic entropy over this pressure range performed in Ref.\,\cite{Wolf2023}, provide first indications of a phase boundary of first order separating AFM order from the VF regime, consistent with the proposed phase diagram in Fig.\,\ref{fig:Fig1new}\,B. To clarify the character of this phase boundary, we performed isothermal measurements of relative length changes as a function of pressure, as illustrated in Fig.\,\ref{fig:Fig2}\,B and \ref{fig:Fig2}\,C. Starting at an elevated temperature of  $100$\,K, where no significant effects of charge and spin fluctuations are expected, $\Delta L_a(T=\mathrm{const.},p)/L_a$ decreases approximately linearly with pressure (orange curve), reflecting a normal elastic response, in accordance with Hooke’s law of elasticity. However, for data sets taken at temperatures between 60\,K and 47\,K (light green curves), we find distinct nonlinearities, that is, a rising change in the slope, indicating an increasing softening of the lattice when entering the VF regime with increasing pressure. Upon further cooling to 42\,K (dark green curves), the data reveal a kink around 14\,MPa, followed by a broad s-shaped anomaly. At 40\,K, first signs of jump-like anomalies are observed, which grow in size on cooling to 32\,K (Fig. \ref{fig:Fig2}\,C, see also Fig.\,S10). The discontinuous length changes, together with the hysteresis revealed by increasing and decreasing the pressure (Fig.\,S11\,A), provide clear evidence of the first-order character of the pressure-induced transition below about 40\,K; see also the discontinuous changes observed in the resistivity (Fig.\,S11\,B) and the magnetic moment (Fig.\,S11\,C). 
\begin{figure*}[t!]
\centering
	\includegraphics[width=0.9\textwidth]{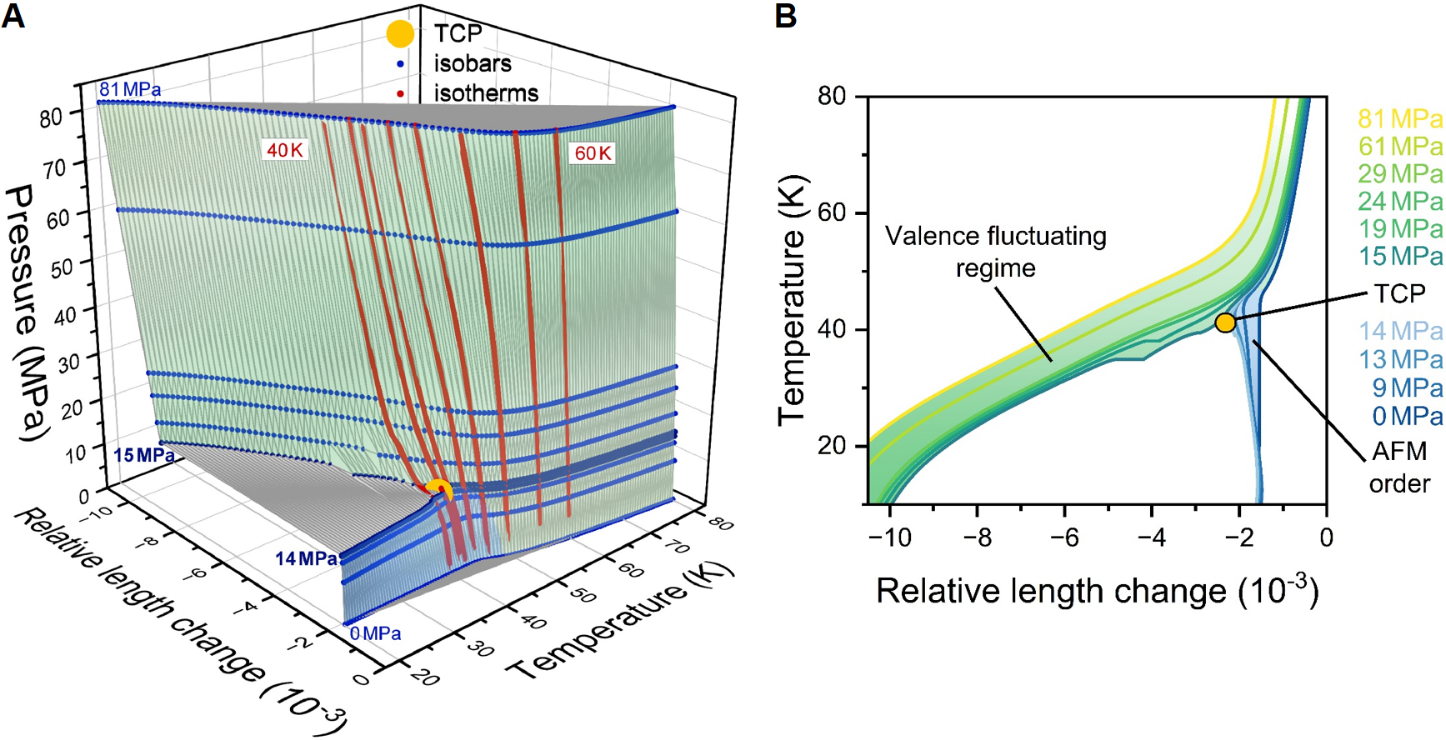}
	\caption{\textbf{Experimentally determined phase diagram of EuPd$_{\mathbf{2}}$(Si$_{\mathbf{0.89}}$Ge$_{\mathbf{0.11}}$)$_\mathbf{2}$ with a tricritical point.} (\textbf{A}) Three-dimensional representation of the relative length change, $\Delta L_a/L_a$, measured along the $a$ axis upon variation of temperature $T$ at $p$ = const.\,(blue data points) and as a function of pressure $p$ at $T$ = const.\,(red data points) of sample \#2a. The data were normalized to their values at $0.1$\,MPa and $100$\,K. The surface colored in green (blue) represents regions where valence fluctuations (antiferromagnetic order) dominate. (\textbf{B}) Projection of data taken at $p$ = const. conditions on the ($T, \Delta L/L$) plane. Areas colored in green (blue) correspond to regions where valence fluctuations (antiferromagnetic order) dominate. The yellow-filled circle in (A) and (B) marks the point where bifurcation is observed in $\Delta L_a/L_a$ and is assigned to the tricritical point ($p_\mathrm{TCP}, T_\mathrm{TCP}$) = $(14\pm1\,\mathrm{MPa}, 41\pm1\,\mathrm{K})$.}
	\label{fig:Fig3}
\end{figure*}

The data in Fig.\,\ref{fig:Fig2} emphasize the particularly high sensitivity of $\Delta L_a(T)/L_a$ to discriminate AFM order (small variations in $\Delta L_a(T)/L_a$) from VF (strong variations in $\Delta L_a(T)/L_a$). 
This observation can be used in Fig.\,\ref{fig:Fig3}\,A, which compiles $\Delta L_a(p,T)/L_a$ data obtained under isothermal (red curves) and isobaric (blue curves) conditions, to identify phase-space regions, indicated by colored surfaces, where AFM order prevails (blue surface) and those regions governed by strong VF (green surface) -- an even finer differentiation is possible by considering the pressure response of $\Delta L_a/L_a$, see the discussion below. The gray area in Fig.\,\ref{fig:Fig3}\,A, spanned by the isobars at 14 and 15\,MPa, corresponds to the range where the transition between AFM order and the VF regime becomes first order. A projection of the data taken at $p$ = const. conditions on the $(\Delta L_a/L_a,T)$ plane is shown in Fig.\,\ref{fig:Fig3}\,B. The figure highlights a point of bifurcation, where, below a critical temperature and depending on pressure, the system shows clear signatures of phase separation. As we argue below, this point of bifurcation, which marks the endpoint of the line of first-order phase transitions, will be used as an estimate of the location of the tricritical point (yellow-filled circle in Figs.\,\ref{fig:Fig3}(A) and (B)), resulting in ($p_{\rm TCP}$, $T_{\rm TCP}$) = $(14\pm1\,\mathrm{MPa},  41\pm1\,\mathrm{K})$.  


\begin{figure}[t!]
\centering
	\includegraphics[width=.5\linewidth]{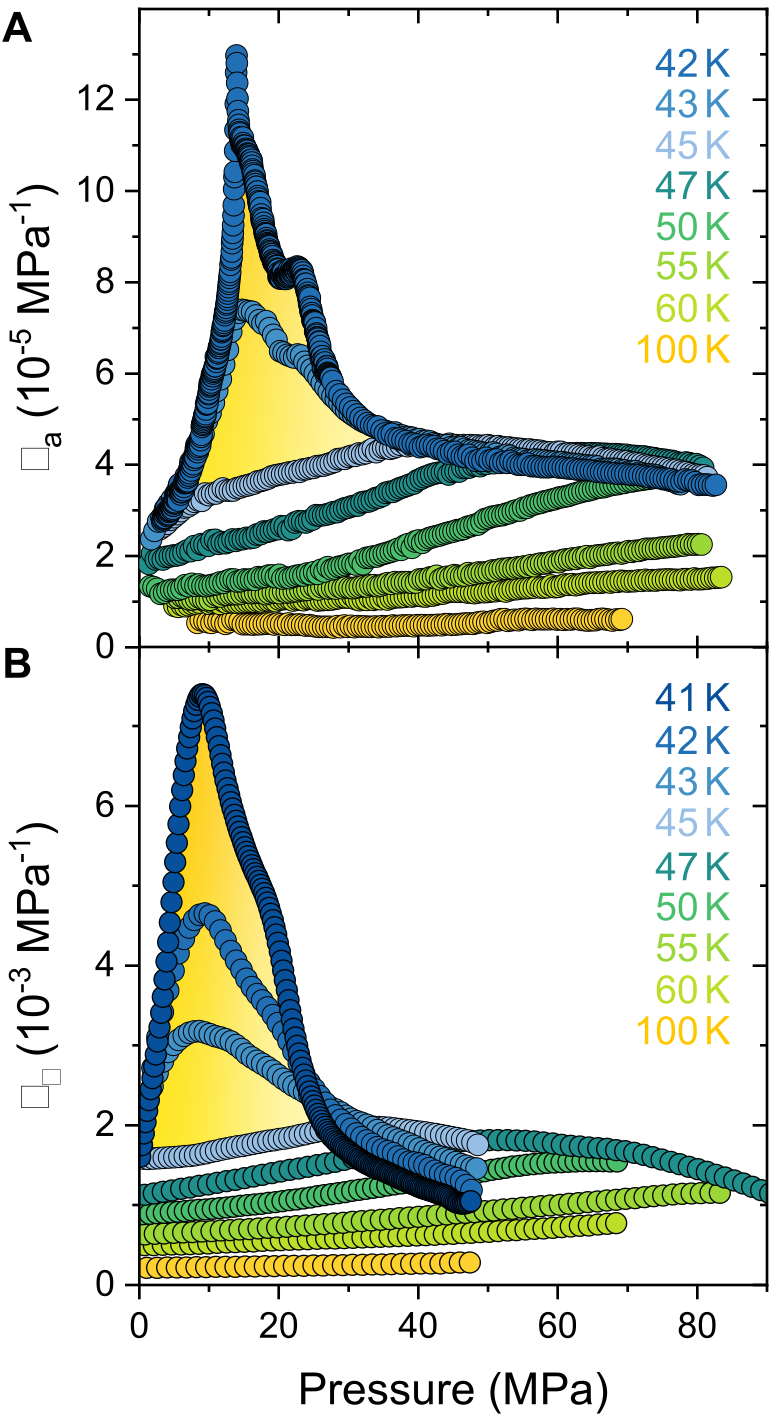}
	\caption{\textbf{Effect of pressure on approaching the tricritical point.} Compilation of experimental data on \EuPdSiGex\ derived from pressure sweeps at selected fixed temperatures. (\textbf{A}) Linear isothermal $a$-axis compressibility, $\kappa_a$ = $-1/L_a \cdot \left( d L_a/d p \right )_T$, of sample \#2a. (\textbf{B}) Pressure derivative of the in-plane electrical resistivity, $\kappa_{\rho}$ = $1/ \rho \cdot \left( d \rho/d p \right )_T$, of sample \#2a\_2, a fragment of \#2a, for the same selected temperatures as in (A) using the same color code. The differentials in (\textbf{A}) and (\textbf{B}) were approximated by calculating the difference quotients. Areas highlighted in yellow indicate the strong increase in the pressure response revealed upon approaching the tricritical point. The two-peak feature can be attributed to inhomogeneities in the sample, which are presumably due to a slightly different Ge ratio with two accumulation points corresponding to two (majority and minority) volume fractions (see Sec. B in SI Appendix for details).}
	\label{fig:Fig4}
\end{figure}
The largely different strain-stress response observed along the $a$ axis in reducing the temperature from 100\,K to 42\,K and finally 32\,K, illustrated in Fig.\,\ref{fig:Fig2}\,B and \ref{fig:Fig2}\,C, reflects drastic changes in the elastic behavior of the material within this temperature window. For a deeper inspection, we consider in Fig.\,\ref{fig:Fig4}\,A the linear isothermal $a$ axis compressibility $\kappa_a$ = $-1/L_a \cdot \left( d L_a/d p \right )_T$. At 100\,K and in the pressure range investigated here, the system is in a paramagnetic state with a slightly reduced Eu moment \cite{Wolf2023}. In the absence of noticeable charge and magnetic fluctuations here, $\kappa_a$($p$, $T$ = 100\,K) (yellow spheres) is small $\sim5\times10^{-6}$\, MPa$^{-1}$ (consistent with results reported in Ref.\,\cite{Batlogg1982}) and featureless. 

Significant changes in the elastic behavior are apparent in the data taken at 60\,K, and these changes become increasingly more pronounced when the temperature is reduced further to 45\,K (green spheres in Fig.\,\ref{fig:Fig4}\,A), i.e., by penetrating deeper into the VF regime. The anomalous elastic behavior manifests itself in two respects. First, there is an overall growth in $\kappa_a$ up to about one order of magnitude with decreasing temperature, a trend that also includes the behavior down to vanishingly small pressures. Second, $\kappa_a$ increases with increasing pressure, with the appearance of a broad maximum that shifts to lower pressures as the temperature decreases. Based on its position in the $(p,T)$ phase diagram (cf. Figs. \ref{fig:Fig5} and S9), the maximum in $\kappa_a(p,T=\mathrm{const.}$) can be assigned to the valence-crossover line $T'_\mathrm{V}(p)$.

A markedly different behavior in $\kappa_a(p,T=\mathrm{const.})$ is revealed when the temperature is slightly reduced further, first to 43\,K and then to 42\,K (blue spheres), by the formation of a narrow, sharp maximum whose height increases drastically. The shape of the anomaly observed here, which reveals a rapid drop on its low-pressure flank, indicates a pressure-induced phase change. Considering the absence of discontinuities in the corresponding $ \Delta L_a (p, T = \mathrm{const.})/ L_a$ vs. $p$ data (Fig.\,S10), which lack any hysteresis in increasing and decreasing pressure, we conclude that this phase change at 43\,K and 42\,K is of second order. The finding of a pronounced maximum in $\kappa_a$, corresponding to an extraordinarily softening of the in-plane lattice within a narrow $T$-$p$ range (highlighted by the yellow area), marks the major result of this work. As we argue below, we consider this effect to reflect \textit{elastic tricriticality}, i.e., a critical enhancement of the compressibility when tuning the system to its TCP. Note that, for a tetragonal system, the volumetric compressibility is given by $\kappa_V = 2 \kappa_a + \kappa_c$, where $\kappa_c$ is the corresponding $c$-axis compressiblity. Consequently, a divergence of $\kappa_a$ implies a divergence of $\kappa_V$. We like to point out that, in addition to the main maximum in $\kappa_a$ at around 14\,MPa, a second, less pronounced maximum appears at around 23\,MPa in the data sets at 43\,K and 42\,K. We assign this feature to sample inhomogeneities, i.e., a minor volume fraction with a slightly different Ge concentration resulting in a somewhat enhanced critical pressure (see Sec. B and Fig.\,S9 in SI Appendix).

It is interesting to compare the anomalous elastic properties with the corresponding response of the resistivity to pressure changes by considering the quantity $\kappa_{\rho}$ = $1/ \rho \cdot \left( d \rho/d p \right )_T$. The data depicted in Figure \ref{fig:Fig4}\,(B) have been obtained by calculating the difference quotient of $\Delta \rho (p, T = \mathrm{const.}) /\rho$ vs. $p$ data (see Fig.\,S13). 
In fact, the $\kappa_{\rho}$ data reveal a strikingly similar behavior to the elastic response (Fig.\,\ref{fig:Fig4}\,A) including a small and featureless $\kappa_{\rho}(p, T=\mathrm{const.})$ at 100\,K that grows in size as the temperature decreases to 60\,K, and the appearance of a broad maximum for the data at 47\,K and 45\,K, which is shifted to lower pressures with decreasing temperature. Most importantly, we find the same drastic change in $\kappa_{\rho}(p,T=\mathrm{const.})$ when the temperature is reduced further to 43\,K, where a pronounced narrow maximum is formed, which is asymmetric in shape with a steep low-pressure flank, and the height of which increases strongly as the temperature is reduced to 42\,K and 41\,K. As for $\kappa_a$, the $\kappa_{\rho}$ data also reveal a small and less pronounced feature on the high-pressure flank of the main maximum. 

\section*{Discussion}



\subsection*{Phase diagram and tricritical point of \EuPdSiGex} \label{phase diagram}

\begin{figure*}[t!]
\centering
	\includegraphics[width=0.67\textwidth]{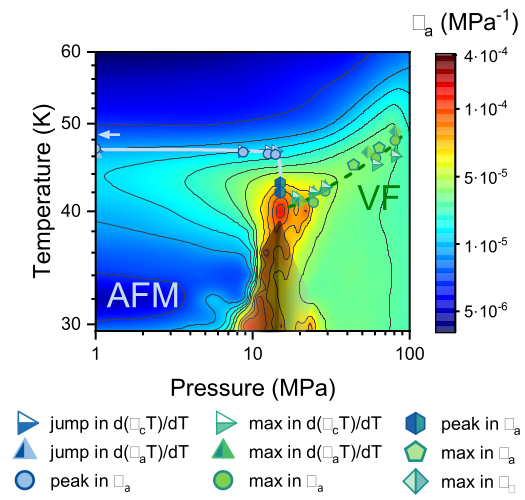}
	\caption{\textbf{Phase diagram and in-plane compressibility of EuPd$_{\mathbf{2}}$(Si$_{\mathbf{0.89}}$Ge$_{\mathbf{0.11}}$)$_\mathbf{2}$ in the vicinity of its tricritical point.} Blue data points correspond to second-order phase transition anomalies at the formation of antiferromagnetic order at $T_{\mathrm{N}}(p)$ (white line), including (i) a jump-like anomaly in $d(\chi \cdot T)/dT$, where $\chi$ is the magnetic susceptibility measured in a magnetic field of $H$ = 10\,kOe, aligned along ($\chi_a$) and perpendicular ($\chi_c$) to the in-plane $a$ axis, yielding an easy-plane anisotropy (see Fig.\,S5\,A and Ref.\,\cite{Wolf2023}), and (ii) a pronounced peak in the coefficient of thermal expansion, $\alpha_a = L_a^{-1} (dL_a/dT)$ (Fig.\,S5\,B). Green symbols correspond to the positions of broad maxima associated with valence-crossover behavior at $T'_\mathrm{V}(p)$ (green dashed line) revealed in $d(\chi \cdot T)/dT$, in $\alpha_a$, in the in-plane compressibility $\kappa_{\mathrm{a}}$ and  in the pressure derivative of the in-plane resistivity $\kappa_\rho$. The black semi-transparent line marks the first-order phase transition  separating AFM order at low pressure from a regime of strong VF at higher pressures. The black shaded area indicates the regime where hysteresis effects were observed in $p$ sweeps; see Fig.\,S9 for a corresponding diagram which includes also data for the minor volume fraction with slightly enhanced critical pressure. The color plot shows the anomalous contribution to the in-plane compressibility $\kappa_a$ obtained after subtracting the (non-critical) background contribution determined at $T$ = 100\,K.}
	\label{fig:Fig5}
\end{figure*}

Figure \ref{fig:Fig5} summarizes in a $(p,T)$ phase diagram the positions of the various phase transitions and crossover anomalies of \EuPdSiGex\,revealed by measurements of expansivity (cf. Figs.\,\ref{fig:Fig3}, S7, and S10), magnetic moment (cf. Figs.\,\ref{fig:Fig2}\,A, S6, and S8), and resistivity (cf. Figs. \ref{fig:Fig2}\,A, \ref{fig:Fig4}\,B, S12, and S13). 
The figure also displays the variation of the $a$-axis compressibility, $\kappa_a$, in a colored contour plot. On the low-pressure side, the diagram shows a range of AFM order below the N\'{e}el temperature $T_{\mathrm{N}} \approx 47$\,K (blue symbols connected by the white line), which remains practically constant for $p \lesssim$ 12\,MPa (see Figs.\,S5 and S7 for the determination of $T_{\mathrm{N}}(p)$), consistent with results in Ref.\,\cite{Wolf2023}. Apart from the region surrounding $T_{\mathrm{N}}$, the magnetically ordered phase is characterized by a relatively small $\kappa_a$ (deep blue area), of size $\kappa_a\lesssim1\cdot10^{-5}$\,MPa$^{-1}$. This value is about twice as large as that revealed for EuPd$_{{2}}$(Si$_{{0.89}}$Ge$_{{0.11}}$)$_{2}$ at elevated temperatures (100\,K) and found in other Eu-containing 122 systems with non-fluctuating (stable) Eu valence \cite{Hesse1997}. With further increasing pressure, $T_{\mathrm{N}}$ begins to decrease while simultaneously losing its clear signature in a temperature-dependent experiment (see Fig.\,S7A). The reason for this is likely two-fold. First, it may indicate a rapid drop in $T_{\mathrm{N}}(p)$ in an almost vertical manner, making temperature-dependent studies insensitive. Second, on decreasing the temperature in this part of the phase diagram, there is a large and even further growing contribution from VF surrounding the valence crossover (area with a green-to-yellow color change), which may obscure any signature of AFM order, even in pressure-dependent experiments. However, upon further cooling to below 45\,K\,\textemdash\,that is, entering the region where $\kappa_{\mathrm{a}}$ becomes strongly enhanced (red area)\,\textemdash\,the data at 43\,K and 42\,K in fact do reveal sharp peaks as a function of pressure (blue-green hexagons), reflecting a second-order phase transition. We assign these features to a pressure-induced phase transition associated with $T_{\mathrm{N}}(p)$; see the occurrence of magnetic anisotropy in Fig.\,S8\,A, in support of this assignment.

On the high-pressure side of the phase diagram in Fig.\,\ref{fig:Fig5}, VF prevail manifesting themselves in broadened maxima in the temperature and pressure dependence of various quantities (green symbols), which can be assigned to the crossover temperature $T'_\mathrm{V}(p)$ (green dashed line, and see Figs.\,\ref{fig:Fig4}, S6, S7 and S12). As a function of $p$, this crossover temperature increases with a slope $\mathrm{d}T'_{\mathrm{V}}/\mathrm{d}p = (90$\,\textendash\,$100)$\,K/GPa, consistent with previous observations in Ref.\,\cite{Wolf2023}. The regime governed by VF (green area) is characterized by a significantly, i.e., by an order of magnitude enhanced $\kappa_a \approx 5 \cdot 10^{-5}$\,MPa$^{-1}$, making this quantity a highly sensitive thermodynamic probe for the detection of even weak signatures of VF throughout the phase diagram. For temperatures $T\lesssim40$\,K, the phase diagram shows a line of first-order phase transitions (black, semi-transparent line), as evidenced by discontinuities and hysteretic behavior (black, shaded regime surrounding the black line) in various, pressure-dependent probes, notably $\Delta L_a/L_a$, studied under isothermal conditions (see Figs.\,\ref{fig:Fig2} and \ref{fig:Fig3}, S10, and S11). The phase diagram highlights a tricritical point at ($p_{\rm TCP}$, $T_{\rm TCP}$) = $(14\pm1\,\mathrm{MPa},  41\pm1\,\mathrm{K})$ as a consequence of the second-order transition line $T_{\mathrm{N}}(p)$ turning first order upon merging with the crossover line $T'_{\mathrm{V}}(p)$, consistent with the proposed phase diagram in Fig.\,\ref{fig:Fig1new}\,B. 

While key aspects of the phase diagram shown in Fig.\,\ref{fig:Fig5} and its implications for a variety of strongly correlated electron systems based on Ce \cite{Park2006, Park2008a, Scheerer2017} and Eu \cite{Fukuda2003, Hossain2004, Gouchi2020, Ahmida2020a} have been previously discussed \cite{Watanabe2010, Watanabe2011}, we would like to emphasize that clear experimental evidence of a first-order phase transition including a TCP has been missing thus far. Therefore, the present study on EuPd$_{{2}}$(Si$_{{0.89}}$Ge$_{{0.11}}$)$_{2}$ provides a prime example of the proposed VF-induced collapse of AFM order, leading to a TCP, which enables exploration of the interplay of strong fluctuations in charge, lattice, and spin degrees of freedom.

\subsection*{Universality of elastic tricriticality}
\label{subsec:UniversalityElaTric}

The sensitivity of the phase diagram in Fig.\,\ref{fig:Fig5} with respect to hydrostatic pressure already implies a strong coupling of magnetic and valence fluctuations to the crystal lattice of \EuPdSiGex. In the following, we argue theoretically that this elastic coupling qualitatively modifies the tricritical behavior and changes its universality class, resulting in elastic tricriticality. The key signature of elastic tricriticality is a vanishing elastic bulk modulus or, equivalently, a diverging compressibility.

The focus of our interest is the component $\varepsilon$ of the strain tensor that transforms as a volumetric singlet of the irreducible representations of the crystal group. This singlet will  generically couple to the energy density $\phi^2$ of the order parameter $\phi$ associated with the symmetry breaking at a phase transition,
\begin{align} \label{Eq:PhiEpsilonInt}
\mathcal{L}_{\phi-\varepsilon} = \lambda\, \phi^{2} \varepsilon,
\end{align}
where $\lambda$ is the elastic coupling. Whether this coupling 
influences the critical behavior of a continuous phase transition in a fundamental manner follows from arguments given by Rice \cite{Rice1954}. In second-order perturbation theory in $\lambda$, the coupling leads to a correction to the bulk modulus $\delta K \sim - \lambda^2 \langle \phi^2 \phi^2 \rangle$ that is proportional to the energy density autocorrelation function of the order parameter and thus proportional to the critical part of the specific heat $c_{\rm cr}$. As the phase transition is approached, e.g., as a function of temperature $c_{\rm cr} \sim |T-T_c|^{-\alpha}$ varies with the specific heat exponent $\alpha$. 
For a positive $\alpha > 0$, this negative correction to the bulk modulus diverges, i.e., the total bulk modulus vanishes before the critical temperature is reached. The elastic coupling $\lambda$ thus becomes non-perturbative and it can be shown that the crystal indeed becomes elastically unstable \cite{Rice1954}. A non-perturbative elastic coupling necessarily occurs only for certain universality classes of second-order critical points, e.g., the Ising universality class in spatial dimensions $d = 3$ where elastic criticality will eventually prevail. In contrast, for the Heisenberg universality class in $d=3$ the specific heat exponent is negative, and the elastic coupling remains perturbative for small $\lambda$.

However, at the TCP in $d=3$ the specific heat exponent of the theory at zeroth order in $\lambda$ is given by the Gaussian value $\alpha = 1/2$ \cite{Riedel1972}. The elastic coupling at tricriticality is thus always non-perturbative. As a consequence, the elastic coupling eventually induces a crossover to a regime where the tricriticality is strongly coupled to elasticity. Technically speaking, in the framework of the renormalization group the elastic coupling destabilizes the Gaussian tricritical fixed point and gives rise to an elastic tricritical fixed point that governs the tricritical behavior and reflects the change of the universality class \cite{Wegner1974}, see Sec. I and Fig. S14 in the SI for details. At elastic tricriticality, the compressibility diverges with a characteristic critical exponent.

At the TCP of \EuPdSiGex\, the compressibility is indeed strongly enhanced by about a factor of 20 as compared to its value at higher temperatures, see Fig.~\ref{fig:Fig5}. This is strong evidence that elastic tricriticality prevails in \EuPdSiGex\,sufficiently close to its TCP. Although the observations in Fig.~\ref{fig:Fig4}\,A are consistent with a divergent compressibility at the TCP, our data do not yet allow for a reliable extraction of the associated critical exponent.

\subsection*{Charge transport close to elastic tricriticality}
We now turn to the resistivity and discuss its pressure dependence. Close to the TCP, its variation with pressure, $\kappa_{\rho}$ = $1/ \rho \cdot \left(d \rho/d p \right )_T$ is strikingly similar to the behavior of the compressibility $\kappa_a$, see Fig.\,\ref{fig:Fig4}\,B. The close resemblance between $\kappa_{\rho}$ and $\kappa_{a}$ close to the TCP suggests that tricritical fluctuations dominate not only the thermodynamic response but also charge transport. 

This can be rationalized following arguments put forward by Fisher and Langer \cite{FisherLanger1968}. Here, one assumes that the transport relaxation rate is dominated by short-range critical fluctuations such that the critical part of the resistivity $\rho_{\rm cr}$ obtains a singular temperature dependence proportional to the energy density of the order parameter $\rho_{\rm cr} \sim \langle \phi^2 \rangle$, which in turn, via the elastic coupling, is also proportional to the critical part of the volume change $\Delta V_{\rm cr} \sim \langle \phi^2 \rangle$. Consequently, close to the TCP, when critical contributions dominate, the pressure derivative of the resistivity will behave similarly to the compressibility, $\frac{1}{\rho}\partial_p \rho \sim \kappa$, which accounts for the experimental observations.
As $\kappa_{\rho}$ and $\kappa_{a}$ are similarly enhanced close to the valence crossover $T'_\mathrm{V}(p)$, it appears that this argument also applies to some extent to the valence fluctuations.

\section*{Summary and Outlook}
In summary, by the application of various experimental probes to \EuPdSiGex\,, especially through measurements of relative length changes under control of temperature and He-gas pressure, a tricritical point (TCP) has been identified that results from the interplay of long-range antiferromagnetic order and strong valence fluctuations. The tunability of the material's ground state with pressure is associated with strong tricritical anomalies in elastic properties, notably a compressibility that at the TCP is strongly enhanced by about a factor of 20 as compared to its value at higher temperature. We argued that this implies a strong coupling between tricritical fluctuations and elastic degrees of freedom resulting in elastic tricriticality, where the tricritical point of the magnetic phase transition coincides with an isostructural elastic instability of the crystal lattice, i.e., a diverging compressibility. In addition, we found that the pressure derivative of the resistivity exhibits a close resemblance to the compressibility, which we interpreted as evidence for short-range critical fluctuations that dominate charge transport close to the TCP \cite{FisherLanger1968}. 

Theoretically speaking, elastic tricriticality is specified by a critical fixed point in the renormalization group sense with a diverging correlation length and characteristic critical exponents, see Sec. I in the SI. This is in contrast to the elastic criticality found at critical endpoints of valence and Mott transitions, which are only mean-field like without a diverging correlation length. This difference is rooted in the distinct coupling to the volumetric strain singlet $\varepsilon$. Whereas for the CEP a bilinear coupling $\varepsilon\, \psi$ between the strain singlet and the Ising order parameter $\psi$ of the CEP is allowed \cite{Zacharias2012}, such a bilinear coupling is symmetry forbidden for the TCP. 
The bilinear coupling preempts criticality, cuts off the correlation length and renders the CEP mean-field like. In contrast, such a preemption is not possible for the TCP because it is still connected to a line of second-order phase transitions even in the presence of the elastic coupling \eqref{Eq:PhiEpsilonInt}. As a result, 
electronic and elastic degrees of freedom are non-perturbatively intertwined  at elastic tricriticality resulting in non-trivial critical scaling behavior.

The arguments for the emergence of elastic tricriticality are very general. Although EuPd$_{2}$(Si$_{{0.89}}$Ge$_{0.11}$)$_{2}$  provides a prime example associated with antiferromagnetic ordering, we expect that elastic tricriticality will also be relevant to numerous other materials. This includes pressure-induced tricritical points in itinerant ferromagnets  \cite{Brando2016, Flouquet2011}, such as UGe$_2$ \cite{Taufour2010}, and LaCrGe$_3$ \cite{Kaluarachchi2017}, as well as in several ferroelectrics as, e.g., in KH$_2$PO$_4$ \cite{Schmidt1976}, SbSI \cite{Peercy1975} and NH$_4$Cl \cite{Weiner1972}. In fact, the results of the relative length change in NH$_4$Cl are consistent with a divergent compressibility when approaching the materials’s TCP \cite{Weiner1972}. It should also extend to tricritical points induced by magnetic field but that are  amenable to pressure tuning like in NbFe$_2$ \cite{Friedemann2017} and FeCl$_2$. In the latter a substantial softening of a longitudinal elastic mode near the TCP was already observed \cite{Gorodetsky1978}. 

In general, whenever a line of first-order phase transitions can be crossed by the application of pressure, and this line terminates at a critical point, the associated criticality will be strongly influenced by elasticity. This applies to critical endpoints, e.g., for the Mott transition \cite{Zacharias2012}, to tricritical points considered here, but also to other higher-order critical points, for example, bicritical points found, e.g., close to spin-flop transitions. Under certain conditions, elasticity will also be non-perturbatively involved in quantum versions of such pressure-tuned higher-order critical points \cite{Chandra2020,Samanta2022,Sarkar2023}, for example, quantum bicritical points \cite{Tokiwa2013} or quantum tricritical points \cite{Friedemann2017}.

\section*{Materials and Methods} \label{Materials and Methods}

\noindent \textbf{Crystal growth and selection of samples} \newline Ge-substituted EuPd$_2$Si$_2$ was grown using the Czochralski method as previously described in Refs.\,\cite{Peters2023, Kliemt2022}. The investigated EuPd$_2$(Si$_{1-x}$Ge$_x$)$_2$ crystal MP807 was obtained from an initial melt with a nominal Ge concentration of $x_{\rm nom}=0.2$. As shown in Ref.\,\cite{Peters2023}, the Ge incorporation rate in the grown sample is about 50\% for this substitution level which leads to an enrichment of Ge in the melt in the course of the growth. All measurements described in the main text were carried out on samples cut from the same large mother crystal MP807, see Table\,S1 and Fig.\,S2 in SI Appendix for details, including crystal structure, an overview of all samples investigated, and sample characterization. \newline

\noindent \textbf{Energy dispersive X-ray spectroscopy} \newline 
Energy dispersive X-ray (EDX) analysis was used to analyze the sample composition and germanium incorporation using a Zeiss DSM940A scanning electron microscope with an EDAX detector. The statistical relative error of the Ge concentration in the EDX measurement is typically $7$--$10$\,\%  (see Tab. I in \cite{Peters2023}), which limits the determination of small Ge concentrations in the sample.  The substitution level $x_{\rm{EDX}}$ was obtained by averaging multiple measurements taken at different positions at a certain distance from the seed of a single crystal.  \newline

\noindent \textbf{Capacitive dilatometry} \newline
Measurements of the relative length change $\Delta L_a/L_a = [L_a(T) - L_a(T_0)]/L_a(T_0)$, $L_a$ being the sample length along the in-plane $a$ axis and $T_0$ a reference temperature, with a resolution $\Delta L/L \leq 10^{-10}$, were carried out using a capacitive dilatometer (Kuechler Innovative Measurement Technology \cite{Kuechler2012}) installed in a CuBe pressure cell (Unipress Equipment Division, Institute of High Pressure Physics, Polish Academy of Science). The pressure cell is connected via a CuBe capillary to a compressor unit with a $30$\,MPa helium gas bottle serving as a gas reservoir. The system enables measurements of relative length changes to be performed in various modes, including temperature sweeps at constant pressure, as well as pressure sweeps at constant temperature; see Refs.\,\cite{Manna2012, Agarmani2022} for details on measurements of relative length changes under variable He-gas pressure. Thermal expansion measurements were performed at a heating/cooling rate of $\pm$ 3\,K/h, ensuring a pressure stability of $\Delta p/p \lesssim 0.5\%$ at $p \leq 30$\,MPa and $\Delta p/p \lesssim 5\%$ at $p > 30$\,MPa. The pressure values given throughout the manuscript are nominal values that reflect the pressure set at approximately $100$\,K. The actual pressures at the base temperature are slightly lower due to the isochoric cooling of the cell. Pressure sweeps were performed at a rate of $\sim \pm3$\,MPa/h, with a temperature stability of $\Delta T/T \lesssim 0.1\%$. The use of a capacitive dilatometer under variable (He-gas) pressure requires the knowledge of the dielectric function of the pressure-transmitting medium $\varepsilon^{\mathrm{He}}_{\mathrm{r}}(T,p)$. To this end, $\varepsilon^{\mathrm{He}}_{\mathrm{r}}(T,p)$ was determined by measuring the change in the capacitance of the dilatometer for a reference material (high-purity (5N), oxygen-free copper of $4.5$\,mm length) along the same $T$-$p$-trajectory at which the $\Delta L(T,p)/L$ data were collected for the samples under investigation, see Ref.\,\cite{Agarmani2022} for details. The capacitive dilatometry implies a small force $\Delta F$ acting on the sample along the measuring direction given by $\Delta F/\Delta C \approx1.1\,\mathrm{N/pF}$, which depends on the adjusted starting capacitance $C_0 = C_{\mathrm{empty}} + \Delta C$, with $C_{\mathrm{empty}}$ the capacitance of the empty cell without a sample. For the present experiment $\Delta C\approx9$\,pF, which for the geometry of the sample used resulted in a uniaxial pressure of $p_a \approx8$\,MPa.  \newline

\noindent \textbf{DC-susceptibility} \newline
The magnetic susceptibility $\chi = m/H$, with $H$ the magnetic field, is derived from measurements of the magnetic moment $\Delta m$ = $m(T) - m(T_0)$, with $T_0$ a reference temperature, using a commercial superconducting quantum interference device (SQUID) magnetometer (Quantum Design, MPMS XL), equipped with a CuBe pressure cell (Unipress Equipment Division, Institute of High Pressure Physics, Polish Academy of Science). The pressure cell is connected by a CuBe capillary to a compressor and a helium gas bottle \cite{Wolf2022}. The pressure is recorded by a digital manometer (Keller AG, Leo 5) with a resolution of $|p| \leq 0.1$\,MPa, similar to the setup used for dilatometric investigations. Typically, a rate of $\pm 1$\,K/min was used for the temperature sweeps, and a rate of $\pm0.5$\,MPa/min was used for the pressure sweeps.
\newline

\noindent \textbf{AC-resistivity} \newline 
The resistivity $\Delta \rho$ = $\rho(T) - \rho(T_0)$, with $T_0$ a reference temperature, was measured in a four-terminal configuration by using an AC-resistance bridge LR700 from Linear Research operating at a frequency of 15.9\,Hz. Gold wires ($25\mu$m diameter) were attached along the $a$ axis of the sample using silver paste. For measurements under hydrostatic He-gas pressure, a CuBe pressure cell connected to a compressor unit (Unipress Equipment Division, Institute of High Pressure Physics, Polish Academy of Science) was used via a CuBe capillary. The pressure is recorded by a digital manometer (WIKA, CPG 1500) with a resolution of $|p| \leq 0.1$\,MPa. Temperature sweeps were performed at $\pm 0.2$\,K/min, and pressure sweeps at $\pm 0.5$\,MPa/min.

\clearpage 

%


\nolinenumbers

%
%
%
%
%
%


\paragraph*{Acknowledgments:}
We would like to thank Ch. Geibel, J. Schmalian and N. Sander for valuable discussions, as well as S. Hartmann for her support in the early stage of the experiments. 
\paragraph*{Funding:}
This work was supported by the Deutsche Forschungsgemeinschaft (DFG, German Research Foundation) through TRR 288 - 422213477 (projects A01, A03, A06 and A11).
\paragraph*{Author contributions:}
Conceptualization: B.W., and M.L. 
Methodology: K.K., C.K, B.W., M.G., and M.L. 
Investigation: Y.A., A.A., J.Z., U.T., M.P., M.O., B.W., and M.L. 
Formal analysis: Y.A., A.A., L.F., J.Z, U.T., M.P., M.O., B.W., M.G., and M.L. 
Resources: Y.A., M.P., M.O., K.K., and C.K. 
Funding acquisition: C.K, B.W., M.G., and M.L. 
Validation: Y.A., A.A., U.T., B.W., and M.L. 
Supervision: C.K, B.W., M.G., and M.L. 
Project administration: B.W. and M.L. 
Writing—original draft: Y.A., B.W., M.G., and M.L. 
Writing—review and editing: Y.A., A.A., L.F., J.Z., U.T., M.P., M.O., K.K., C.K., B.W., M.G. and M.L.
\paragraph*{Competing interests:}
There are no competing interests to declare.
\paragraph*{Data availability:}



The article and its Supporting Information Appendix contain all the data needed to evaluate the conclusions in this study. The processed data underlying the figures in this study are openly accessible in the Goethe University data repository (GUDe).


\subsection*{Supplementary Information}
Supplementary Notes 1 to 8\\
Tables S1 and S2\\
Figures S1 to S15\\
References [68] to [80]


\newpage


\renewcommand{\thefigure}{S\arabic{figure}}
\renewcommand{\thetable}{S\arabic{table}}
\renewcommand{\theequation}{S\arabic{equation}}
\renewcommand{\thepage}{S\arabic{page}}
\setcounter{figure}{0}
\setcounter{table}{0}
\setcounter{equation}{0}
\setcounter{page}{1} 


\begin{center}
    \LARGE \textbf{Supplementary Information for}\\[0.5cm]
    \Large \textbf{\scititle}\\[0.5cm]
    \large Yassine Agarmani$^\ast$,
	Arif Ata,
    Lars Franke,
	Jan Zimmermann,
    Ulrich Tutsch,
    Marius Peters,
    Michelle Ocker,
    Kristin Kliemt,
    Cornelius Krellner,\\
    Bernd Wolf,
    Markus Garst,
    Michael Lang\\[1cm]
\small$^\ast$Corresponding author. Email: agarmani@physik.uni-frankfurt.de
\end{center}

\vspace{0cm}
\tableofcontents
\vspace{1cm}
\hrule
\newpage

\newpage


\section*{Supplementary Notes}
\addcontentsline{toc}{section}{Supplementary Note 1: Crystal structure and samples investigated}
\subsection*{A: Crystal structure and samples investigated}
\begin{figure} [h] 
	\centering
	\includegraphics[width=0.4\textwidth]{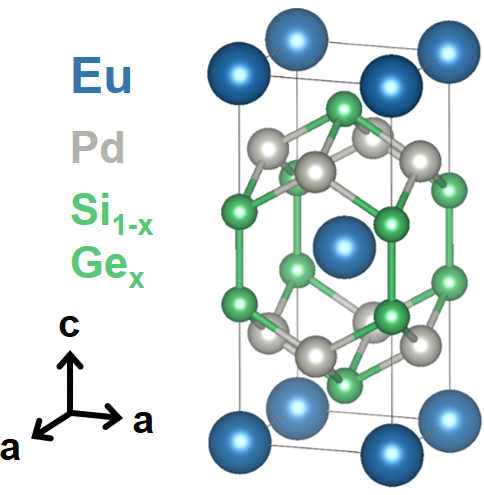} 
	\caption{\textbf{Crystal structure of tetragonal EuPd$_2$(Si$_{1-x}$Ge$_x$)$_2$}. Visualization of the tetragonal ThCr$_2$Si$_2$ (122) structure for the EuPd$_2$(Si$_{1-x}$Ge$_x$)$_2$ system. According to Ref.\,[\citenum{Hoffmann1985}], the structure consists of Eu-layers sandwiched between Pd$_2$[Si(Ge)]$_2$-layers, where beside the covalent Si-Pd bonding also indications for a metal-metal Pd-Pd bonding were found. We believe that the latter, in combination with the strong covalent Si(Ge)-Si(Ge) bonds along the $c$ axis, are responsible for the anisotropic lattice response to changes of the Eu valence. The thick green lines illustrate the strong covalent bonds between the Si(Ge)-Si(Ge) along the $c$ axis.}
	\label{fig:S1} 
\end{figure}

\begin{figure} [t!]
	\centering
	\includegraphics[width=1\textwidth]{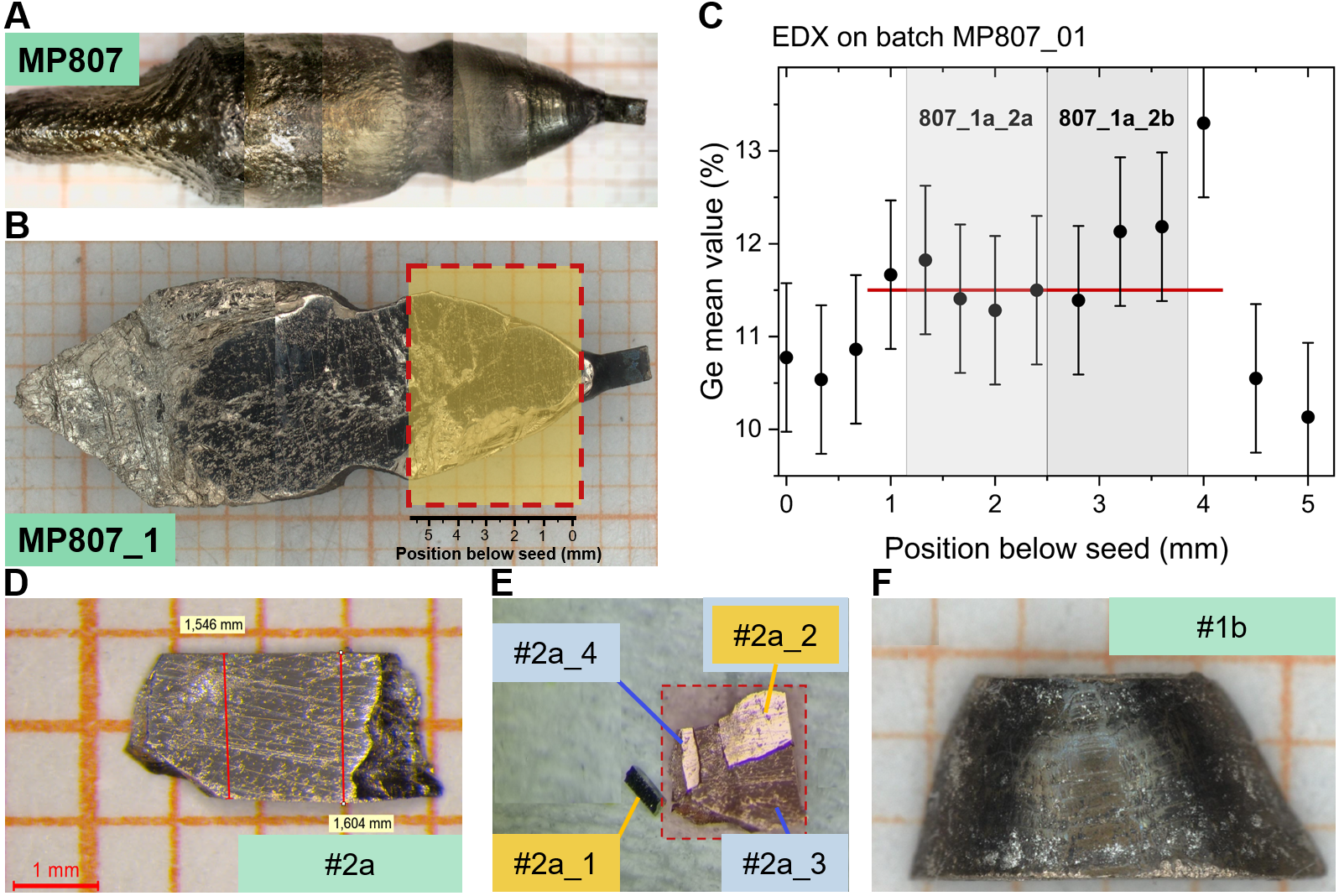} 
	\caption{\textbf{EuPd$_2$(Si$_{1-x}$Ge$_x$)$_2$ single crystal grown by Czochralski technique and EDX characterization.} \textbf{(A)} Result of the Czochralski growth process MP807. \textbf{(B)} Crystal shown in (A) after cut in half. The yellow box with the position axis in mm marks the area where the energy dispersive x-ray (EDX) analysis was carried out. \textbf{(C)}  Mean Ge concentration determined by EDX as a function of distance from the seed position. The EDX analysis for the sample MP807\_1a\_2a used for the measurements of relative length changes reveals $x_{\rm{EDX}}=0.115\pm 0.008$ (red line), i.e., constant within the error bars. Beyond 4\,mm, the Ge concentration drops due to formation of an Eu-rich secondary phase caused by the changes in the melt composition. Gray boxes mark regions where the single-crystalline samples were extracted. \textbf{(D)}-\textbf{(F)} Images of the samples studied with sample numbers.}
	\label{fig:S2} 
\end{figure}

\EuPdSi\, crystallizes in the tetragonal ThCr$_2$Si$_2$ (122) structure \cite{Sampathkumaran1981} shown in Fig.\,\ref{fig:S1}. The samples investigated in this work were all cut from the same mother crystal MP807 (Fig.\,\ref{fig:S2}) to minimize effects related to sample-to-sample variations. A list of the samples and their designation is given in table \ref{tab:S1}. The table also includes related samples studied in Refs.\,\cite{Wolf2023, Wolf2023a}.  \newline
\newline
\begin{table} 
	\centering
	\caption{\textbf{Overview of the EuPd$_2$(Si$_{1-x}$Ge$_x$)$_2$ samples investigated.}
		The sample designation consists of the number of the growth experiment (MP followed by three digits, e.g. MP807). All experiments described in the main text were performed on samples cut from the mother crystal MP807, which was cut into halves (MP807\_1a and MP807\_ 2a), see Fig.\,\ref{fig:S2}\,B. The combination of numbers and letters that follows correspond to the sample number specifying the individual samples used for the various experiments, see Fig.\,\ref{fig:S2}\,D, E, and F. The table allocates which sample was used for the measurements of relative length changes, $\Delta L/L$, the resistivity, $\Delta \rho /\rho$, and the magnetic moment, $\Delta m/m$, described in the main text. The symbol $x${\textdagger} ($x^{*}$) indicates that data were taken in constant magnetic field $H$ with $H$ perpendicular (parallel) to the $c$ axis. The table also includes crystals from other batches the results on which were previously reported in the literature.}
	\label{tab:S1} 
	\begin{tabular}{lcccc} 
		\\
		\hline
		Batch & Abbreviation & & Measurements performed &  \\
		 Sample designation & sample no.  & $\Delta L/L$ & $\Delta \rho/\rho$ & $\Delta m/m$\\
		\hline
		\textbf{MP807\_1a\_2a} & \#2a  & \textbf{$x$} &  & \\
		\quad MP807\_1a\_2a\_1 & \#2a\_1  &  & \textbf{$x$} & \\
		\quad MP807\_1a\_2a\_2 & \#2a\_2  &  & \textbf{$x${\textdagger}} & \textbf{$x$}\\
        \quad MP807\_1a\_2a\_3 & \#2a\_3  &  &  & \textbf{$x^{*}$}\\
		\quad MP807\_1a\_2a\_4 & \#2a\_4  &  &  & \textbf{$x$}\\
		\textbf{MP807\_1a\_1b} & \#1b  & \textbf{$x$} &  & \textbf{$x$}\\
		\textbf{MP401\_3} & \#3  & \textbf{$x$}, \onlinecite{Wolf2023a} &  & \\
        \textbf{MP807\_2a\_1} & \#4  &  &  & \onlinecite{Wolf2023, Wolf2023a} \\
        \textbf{MP805\_1a\_2} & \#5  &  \onlinecite{Wolf2023a} &  &  \\
        \hline
	\end{tabular}
\end{table}

\newpage
\clearpage
\subsection*{B: Sample characterization}
\addcontentsline{toc}{section}{Supplementary Note 2: Sample characterization}
The EuPd$_2$(Si$_{1-x}$Ge$_x$)$_2$ single crystals were characterized by various probes, including energy dispersive X-ray (EDX), powder X-ray diffraction and heat capacity measurements, see Refs.\,[\citenum{Kliemt2022, Peters2023}] for details. Through a combined chemical and structural analysis, it was found that (i) for Czochralski-grown unsubstituted EuPd$_2$Si$_2$ single crystals there is a small variation (of order 1\,\%) of the Pd-Si ratio along the growth direction \cite{Kliemt2022}, and (ii) that a variation of this ratio as small as 1\,\% can cause a significant shift in the valence-crossover temperature $T'_{\mathrm{V}}$ of about 15\,K. Samples with lower $T'_{\mathrm{V}}$ were formed at the beginning of the growth process, while those with higher $T'_{\mathrm{V}}$ were formed later. 
A similar trend was found for the Ge-substituted samples with $x = 0.11$ (see Supplementary Information of Ref.\,\cite{Ocker2026}). The samples used in this study were cut at a distance of $1\,\mathrm{mm} < x <4\,\mathrm{mm}$ below the seed crystal (see Fig.\, S2\,B and C), where the expected shift of $T_\mathrm{V}^{\prime}$ is small. 

The analysis of crystal MP807, Fig. \ref{fig:S2}\,B, revealed a substitution level comparable to that reported previously for $x_{\rm nom}=0.2$ \cite{Peters2023}. Note that the systematic relative error of the EDX measurement is typically $7$--$10$\,\%, which means that the determination of small Ge concentrations is limited by experimental uncertainty. The substitution level in the sample, $x_{\rm{EDX}}= 0.115(8)$, 
Fig.\ref{fig:S2}\,C, was obtained by averaging multiple measurements taken at different positions at a certain distance from the seed of a single crystal. It was found to be constant within the resolution of the measurement in the first $4$\,mm below the seed. 
A more precise determination of the Ge concentration can be achieved using wavelength dispersive X-ray  (WDX) and micro X-ray fluorescence ($\mu$-XRF) spectroscopy as recently demonstrated in Ref.\,\cite{Ocker2026} for a similar Ge-substituted sample. Consistent with the results in Ref.\,\cite{Ocker2026}, the EDX analysis of crystal MP807 also shows the formation of side-phase inclusions at distances greater than $4$\,mm from the seed. In addition, the $\mu$-XRF analysis in Ref.\,\cite{Ocker2026} shows tiny variations in the Ge content in the upper part of the crystal that are below the EDX and WDX detection limits. Such minor variations in the chemical composition are probably responsible for the slight variations in the physical properties discussed below.

\begin{figure}[t!] 
	\centering
	\includegraphics[width=0.5\textwidth]{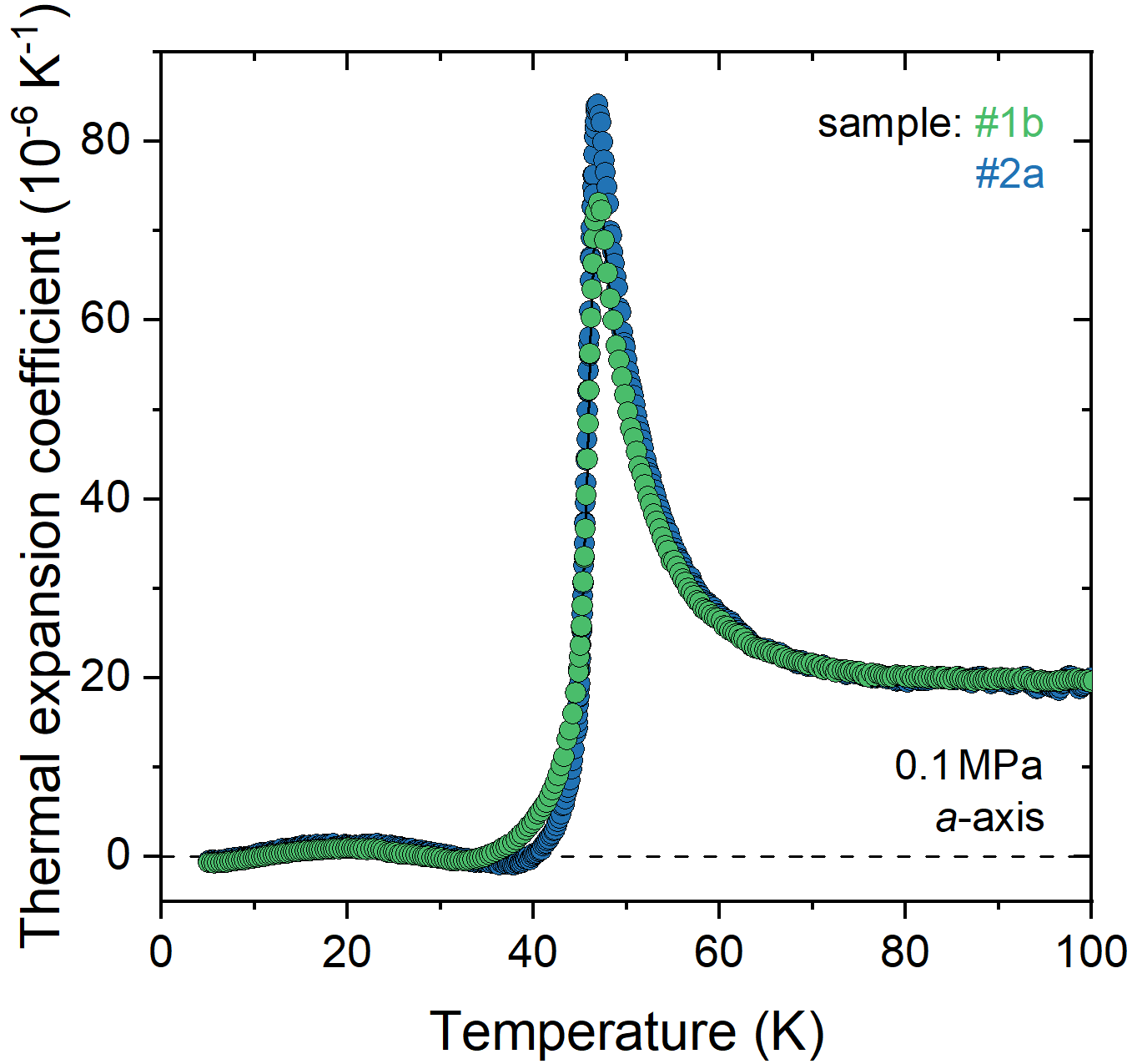} 
	\caption{\textbf{Coefficient of thermal expansion around the temperature of magnetic ordering.} 
		In-plane coefficient of thermal expansion $\alpha_a(T)=L_a^{-1}dL_a/dT$ as a function of temperature at ambient pressure (0.1\,MPa) for sample \#1b (green) and sample \#2a (blue). Whereas the position of the maximum, corresponding to the Néel temperature $T_\mathrm{N}$, is identical for both crystals, there is a small difference between the data sets slightly above $T_\mathrm{N}$, and around 40\,K, where a small negative contribution is observed the size of which is different for both crystals.}
	\label{fig:S3} 
\end{figure}

\textbf{Sample-to-sample variations with respect to $T_\mathrm{N}$ and $T'_{\mathrm{V}}$.}
Despite the small homogeneity range of EuPd$_2$(Si$_{1-x}$Ge$_x$)$_2$, samples from different batches prepared by using the same starting compositions yield consistent results with regard to the magnetic order at ambient pressure (see Fig.\,\ref{fig:S3}), the valence-crossover behavior at finite pressure and the overall phase diagram. More specifically, as discussed in Refs.\,\cite{Wolf2023, Wolf2023a}, samples prepared from the batches MP\,807 and MP\,805, which have the same Ge-concentration $x$ = 0.11 as determined by EDX, show very similar magnetic behavior as a function of temperature at varying hydrostatic pressure. There are only minor differences in the magnetic susceptibility $\chi(T)$ in the ordered magnetic phase at ambient pressure below 25\,K and small differences in the coefficent of thermal expansion $\alpha_a(T)$ slightly below $T_\mathrm{N}$ (see Fig.\,\ref{fig:S3}). Furthermore, there is a tiny difference in the effective magnetic moment $n_{\mathrm{eff}}$ (less than 1\%) in the high-temperature range around 300\,K.
\newline Apart from these small differences, the samples of both batches reveal identical values within the experimental resolution for $T_\mathrm{N}$ = (47.3 $\pm$ 0.2)\,K at $p$ = 0, and identical values for the valence-crossover temperature $T'_{\mathrm{V}}$($p$) for $p > $ 0.1\,GPa (see table I in Ref.\,\cite{Wolf2023}). Measurements under varying pressure consistently reveal large pressure coefficients of $dT'_{\mathrm{V}}/dp$ = +(90 $\pm$ 10)\,K/GPa (MP\,807) and +(100 $\pm$ 10)\,K/GPa (MP\,805) which are practically identical within the margins of error (see Table I in Ref.\,\cite{Wolf2023}).

\textbf{Sample-to-sample variations with respect to $p_\mathrm{TCP}$.} The slight variations in germanium concentration mentioned above can have a significant effect on the absolute critical pressure. This effect is further amplified by the remarkably small pressure scale of this system. This explains the slight deviation in $p_\mathrm{TCP}$ compared to earlier studies \cite{Wolf2023, Wolf2023a} on nominally similar samples. However, these minor chemical variations merely shift the absolute pressure axis without affecting the underlying critical behavior.

\textbf{Indications for sample inhomogeneities.}
Some indications for sample inhomogeneities with regard to the magnetic transition that are not recognizable in temperature-dependent investigations can be revealed in experiments under variable pressure. A closer look at the $\kappa_a(p)$ data in Fig.\,4\,A taken at $T = 43$ and $42$\,K show a dominant, sharp peak around 14\,MPa and a second, distinctly less pronounced maximum around 23\,MPa. 
These data suggest some inhomogeneities, presumably in the Ge concentration \cite{Ocker2026} (and therefore also in the Pd-Si ratio \cite{Kliemt2022}) 
with two accumulation points, corresponding to two (majority and minority) volume fractions, which are characterized by more or less homogeneous but slightly different Ge concentrations and thus slightly different critical pressure values. This notion is corroborated by the observation of two closely spaced discontinuities in $\Delta L_a (T = \mathrm{const.},p) /L_a$ in Fig.\,2\,C and Fig.\,\ref{fig:S10} with the distinctly larger (by a factor of 2) discontinuity at lower pressure, representing the majority phase. As demonstrated in the phase diagram in Fig.\,\ref{fig:S9}, which also includes the smaller jumps in $\Delta L_a (T = \mathrm{const.},p) /L_a$ at higher pressures, assigned to the minority phase, the corresponding hysteresis closes around 23\,MPa. This finding is consistent with the minority phase having a somewhat enhanced critical pressure of $p_\mathrm{TCP}$ = 23\,MPa. \newline
The above findings are based on the thermodynamic quantity $\kappa_a$ (Fig.\,4\,A) which reflects the bulk property of the large crystal \#2a (Fig.\,\ref{fig:S2}\,D). In the transport quantity $\kappa_{\rho}$ (Fig.\,4\,B) measured on crystal \#2a\_2 (Fig.\,\ref{fig:S2}\,E), a fragment of the crystal \#2a, we find small shifts in the positions of the dominant peak and the subdominant anomaly (now in the form of a shoulder), their relative distance and their sharpness. We attribute these observations to the presence of strain fields which we expect to be present in these crystals in different ways depending on their pristine defect structure after preparation and, in particular, depending on their thermal and pressure history: Given the strong and highly anisotropic lattice effects associated with the temperature-induced valence changes, cooling through the first-order transition is (almost) inevitably accompanied by the generation (or further growth) of structural defects of different sizes and thus the formation of strain fields. It is likely that the strain fields in the smaller fragment \#2a\_2, which actually chipped of crystal \#2a after repeated thermal cycling through the first-order transition, differ from those in crystal \#2a in the early stage of the experiments. Given the strong pressure dependence of $T'_{\mathrm{V}}$, these different strain fields will result in some shifts of the anomalies (maximum or shoulder), their relative distance, and their sharpness, consistent with the experimental observations. \newline
Indications for sample inhomogeneities with regard to the valence-crossover behavior can be seen in temperature-dependent investigations (Fig.\,\ref{fig:S6}). Besides the main feature (rounded maximum) in $d(\chi \cdot T)/dT$, $\alpha_a$ and $-\rho^{-1}d\rho/dT$, the first two (thermodynamic) quantities reveal a small feature on the low-temperature flank which is not visible in the transport quantity. These observations indicate that a small fraction of the sample is characterized by a valence-crossover behavior with a somewhat reduced $T'_{\mathrm{V}}$ compared to the majority phase with slightly higher $T'_{\mathrm{V}}$. 
\newpage
\subsection*{C: Lattice effects around the valence crossover in EuPd$_2$Si$_2$}
\addcontentsline{toc}{section}{Supplementary Note 3: Lattice effects around the valence crossover in EuPd$_2$Si$_2$}

\begin{figure} [b!]
	\centering
	\includegraphics[width=0.75\textwidth]{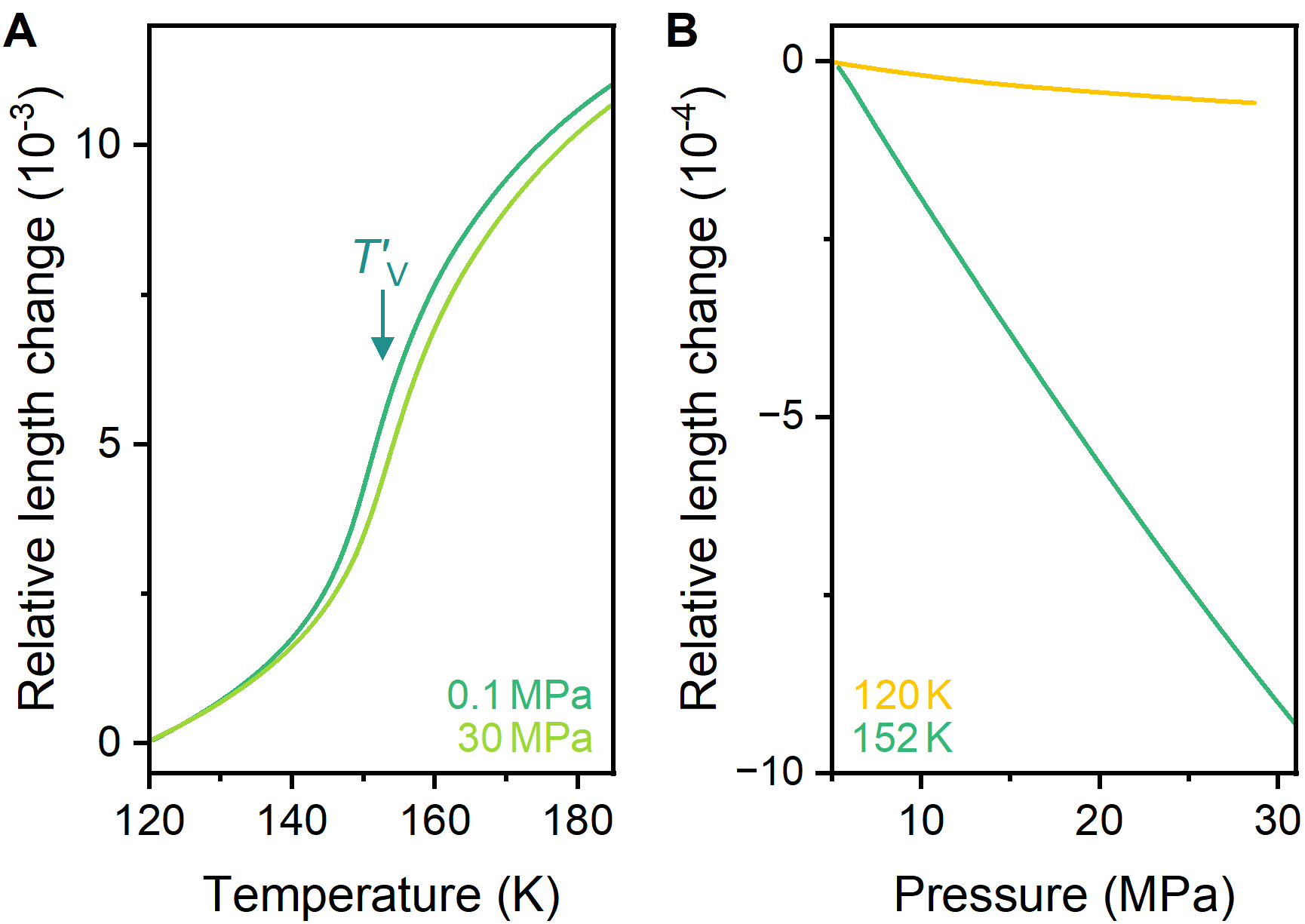} 
	\caption{\textbf{Relative length changes of single crystalline EuPd$_2$Si$_2$.} 
		\textbf{(A)} Relative length changes, $\Delta L_a/L_a$, measured within the tetragonal plane (sample \#3), as a function of temperature at $p$ = 0.1\,MPa (green) and 30\,MPa (light green) around the valence-crossover temperature $T'_{\mathrm{V}}$ \cite{Wolf2023a}. \textbf{(B)} $\Delta L_a/L_a$ as a function of pressure at a temperature $T = 120$\,K, (yellow), i.e., at a temperature distinctly below $T'_{\mathrm{V}}$, where valence fluctuations (VF) are suppressed, and at $T = 152$\,K (green), where VF are strongest.}
	\label{fig:S4} 
\end{figure}

Unsubstituted EuPd$_2$Si$_2$ has been known to reveal a temperature-induced valence change from Eu$^{2.3+}$ around 300\,K to Eu$^{2.8+}$ below about 100\,K \cite{Sampathkumaran1981,Batlogg1982, Kemly1985, Fedchenko2024}. Figure \ref{fig:S4}\,(A) shows the relative length change measured within the tetragonal plane, $\Delta L_a/L_a$, as a function of temperature at nominal ambient pressure $p = 0.1$\,MPa (blue) and at $30$\,MPa (green) of single crystalline EuPd$_2$Si$_2$. A pronounced shrinkage of $\Delta L_a/L_a$ by about a percent is observed as the material is cooled from 170\,K down to 140\,K. In contrast to a first-order transition, generally observed at the valence instability \cite{Segre1982, Wada1999, Wada1999a, Mitsuda2012, Onuki2016}, the $\Delta L_a/L_a$ data for EuPd$_2$Si$_2$ reveal a gradual (yet strong) change consistent with a crossover behavior \cite{Wolf2023a, Peters2023}. By using the point of maximum slope (corresponding to a maximum in $\alpha_a$), results in a crossover-temperature of $T'_{\mathrm{V}}$ = 152\,K (arrow in Fig.\,\ref{fig:S4}\,A). The observed phenomenology indicates that EuPd$_2$Si$_2$ is located on the high-pressure side (crossover region) of the second-order critical endpoint (CEP) terminating the line of first-order valence transitions, see Refs.\,\cite{Wolf2023, Wolf2023a} for further details. Figure \ref{fig:S4}\,(A) also shows $\Delta L_a/L_a$ data taken at $p$ = 30\,MPa, yielding an increase of the crossover temperatures by about $2.5$\,K, corresponding to a pressure dependence of $dT'_{\mathrm{V}}/dp$ = ($83 \pm 10$)\,K/GPa, consistent with results of susceptibility measurements under pressure \cite{Wolf2023}. 
Figure \ref{fig:S4}\,(B) shows $\Delta L_a/L_a$ data as a function of pressure at constant temperature. Data were taken at a temperature of $T = 120$\,K (yellow), which is distinctly below $T'_{\mathrm{V}}$, and at $152$\,K (green), that is, at $T'_{\mathrm{V}}$. The results reveal a significantly higher slope for $T$ = $T'_{\mathrm{V}}$, indicating a particularly soft in-plane lattice within the regime of strong valence fluctuations, compared to the data at 120\,K, where VF are suppressed. 
 
\clearpage

\subsection*{D: Tracing signatures of AFM order and VF from $T$-dependent measurements under $p$ = const. conditions}
\addcontentsline{toc}{section}{Supplementary Note 4: Tracing signatures of AFM order and VF from ...}

\begin{figure} [t!]
	\centering
	\includegraphics[width=0.5\textwidth]{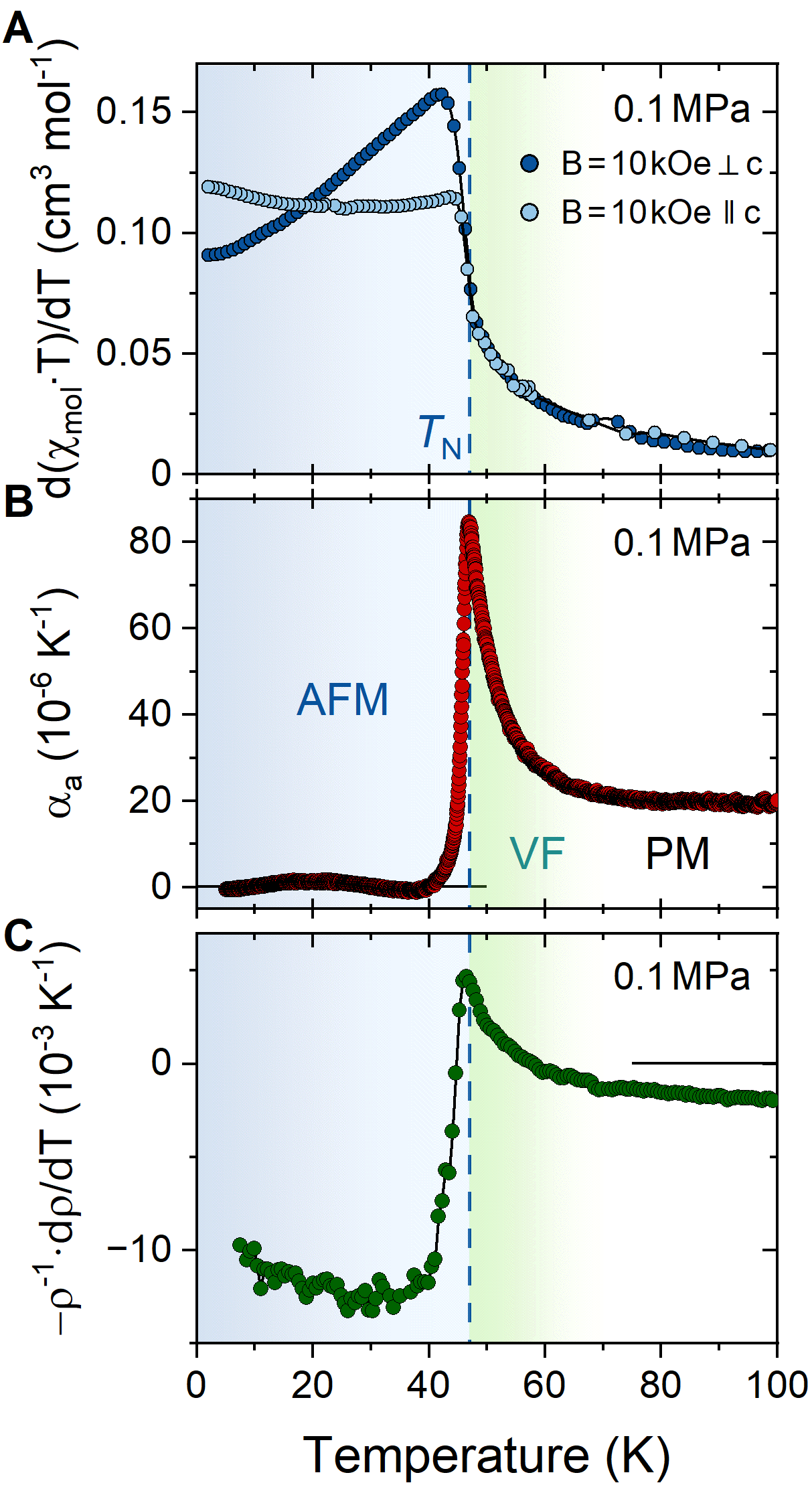} 
	\caption{\textbf{Phase transition into AFM order in \EuPdSiGex\,.} 
		\textbf{(A)} Temperature derivative of the molar magnetic susceptibility $\chi_\mathrm{mol}$ times temperature $T$ at ambient pressure (0.1\,MPa) (samples \#2a\_2 - \#2a\_4), for an external field fo $H$ = 10\,kOe oriented along (light blue circles) and perpendicular to the tetragonal $c$-axis (blue circles). The Néel temperature, $T_{\mathrm{N}}$, indicated by a blue upward arrow, corresponds to the midpoint of the jump-like anomaly in d$(\chi \cdot T)$/d$T$ as described in the text. 
        \textbf{(B)} Coefficient of thermal expansion $\alpha_{a}(T)$ = $1/L_a \cdot \left( d L_a/d T \right )_p$ (red circles) at $p$ = 0.1\,MPa (sample \#2a) on the same temperature scale. \textbf{(C)} Resistivity plotted as $-\rho^{-1} d\rho/dT$ (green circles) at $p$ = 0.1\,MPa (sample \#2a\_2) on the same temperature scale. The green area marks the region with notable valence fluctuations (VF).}
	\label{fig:S5} 
\end{figure}

\begin{figure} [b!] 
	\centering
	\includegraphics[width=0.5\textwidth]{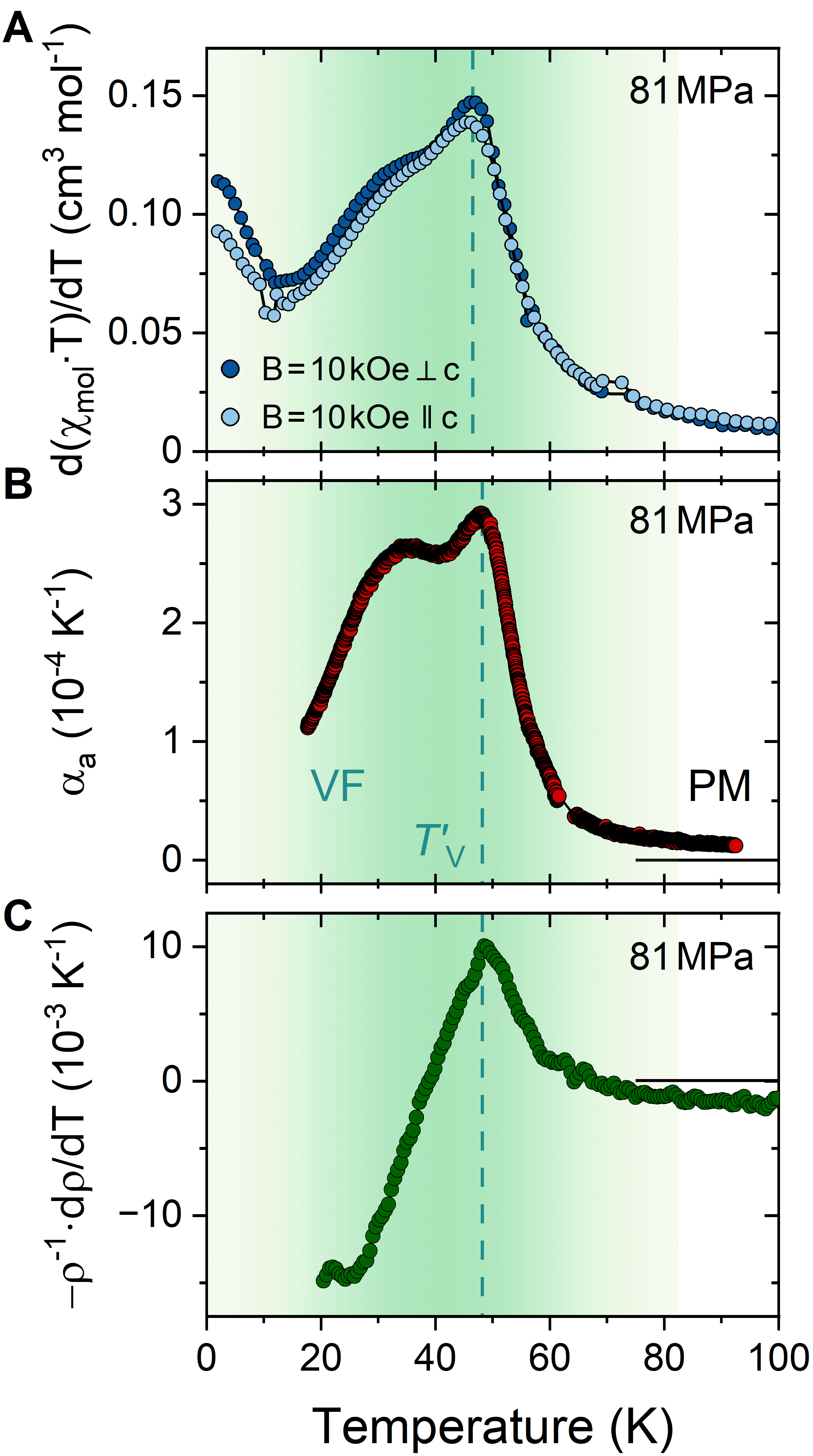} 
	\caption{\textbf{Valence-crossover behavior in \EuPdSiGex\,} 
		\textbf{(A)} Temperature derivative $d(\chi \cdot T )/dT$ at $p$ = 81\,MPa (samples \#2a\_2 - \#2a\_4), for an external field of $H$ = 10\,kOe oriented along  (light blue circles) and perpendicular to the tetragonal $c$-axis (dark blue circles). The valence-crossover temperature, $T'_{\mathrm{V}}$ (vertical dashed line), corresponds to the position of the maximum in $d(\chi \cdot T)/dT$ as described in the text. \textbf{(B)} Coefficient of thermal expansion $\alpha_{a}(T)$ = $1/L_a \cdot \left( d L_a/d T \right )_p$ (red circles) at $p$ = 81\,MPa (sample \#2a) on the same temperature scale. \textbf{(C)} Resistivity plotted as $-\rho^{-1} d\rho/dT$ (light green circles) at $p$ = 81\,MPa (sample \#2a\_2) on the same temperature scale. The green area marks the region around $T'_{\mathrm{V}}$ where valence fluctuations (VF) are strongest. }  
	\label{fig:S6} 
\end{figure}

\begin{figure} [b!] 
	\centering
	\includegraphics[width=0.95\textwidth]{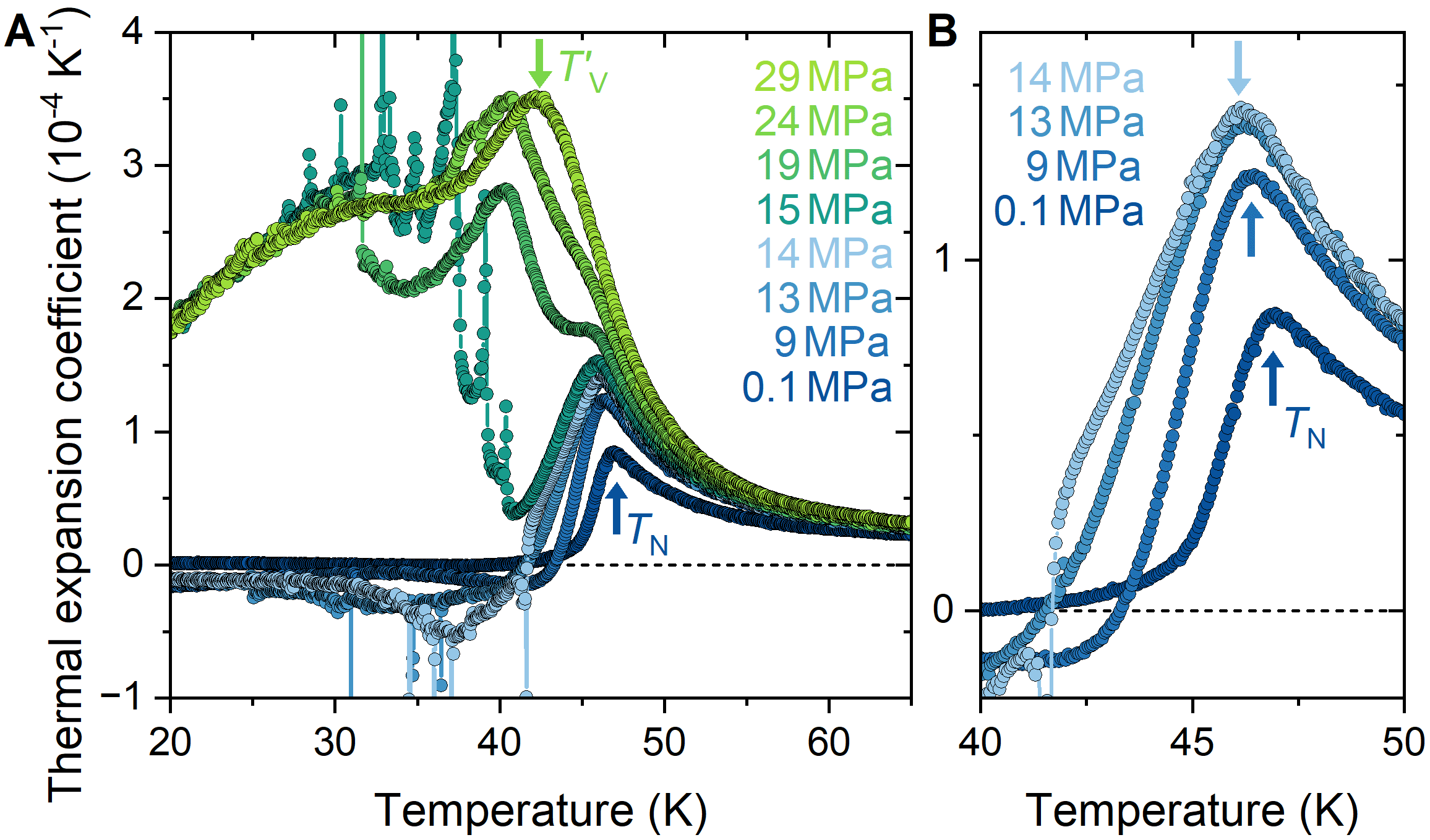} 
	\caption{\textbf{Competition of AFM order and VF in \EuPdSiGex\,} 
		\textbf{(A)} Coefficient of thermal expansion $\alpha_{a}(T)$ = $1/L_a \cdot \left( d L_a/d T \right )_p$  under $p$ = const. conditions for varying pressures 0.1\,MPa $\leq p \leq$ 29\,MPa (sample \#2a). Panel (\textbf{B}) shows a section of the data for $p \leq$ 14\,MPa around the Néel temperature $T_\mathrm{N}$ marked by an arrow.  } 
        \label{fig:S7} 
\end{figure}

\begin{figure} [b!] 
	\centering
	\includegraphics[width=0.85\textwidth]{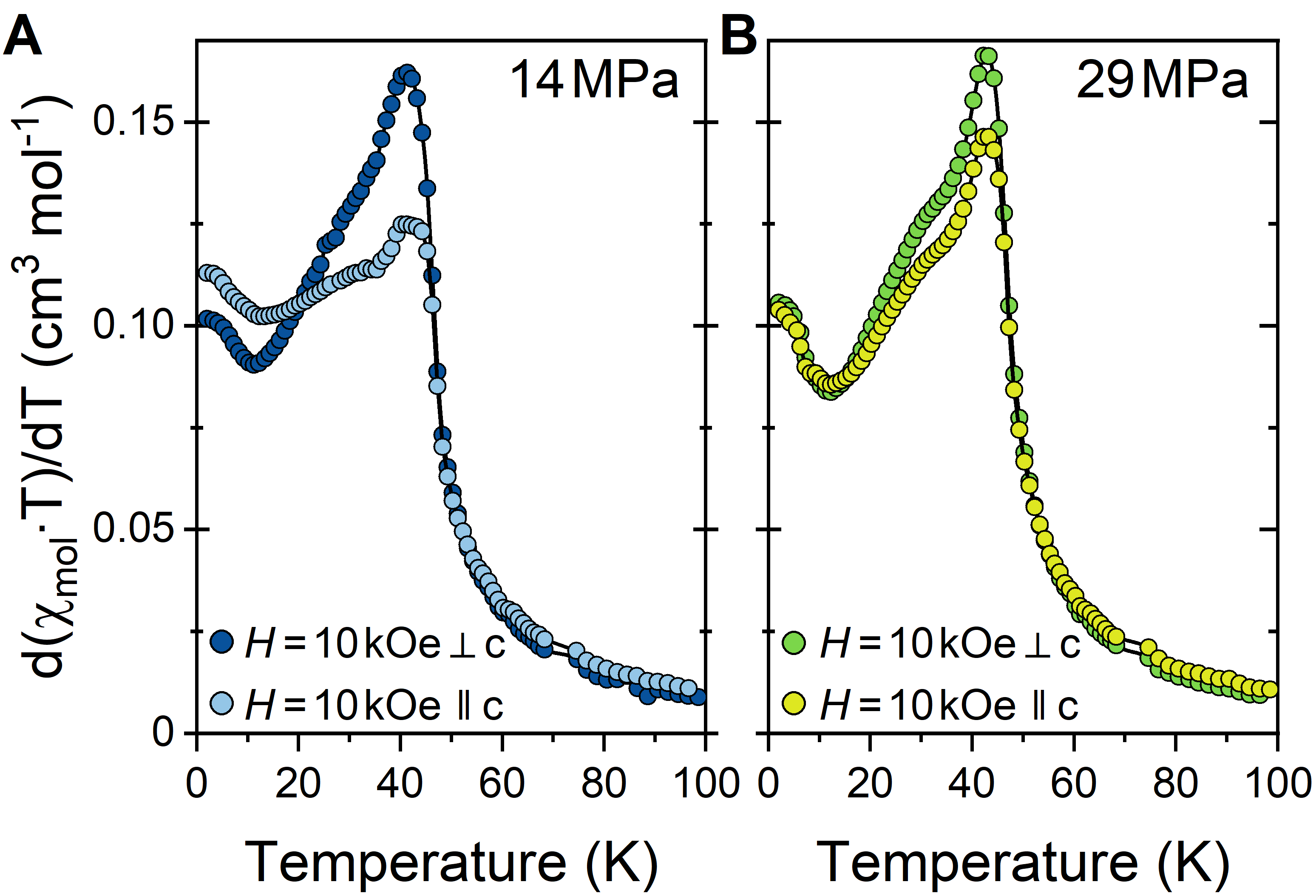} 
	\caption{\textbf{Magnetic signatures of AFM order and VF in \EuPdSiGex\,.} Magnetic susceptibility plotted as $d(\chi \cdot T)/dT$ for $T <$ 100\,K at $p$ = 14\,MPa (A) and $p$ = 29\,MPa (B) for an external field $H$ = 10\,kOe\, oriented along and perpendicular to the tetragonal $c$ axis (samples \#2a\_2 - \#2a\_4).} 
        \label{fig:S8} 
\end{figure}

\textbf{AFM order at $p$ = 0.1\,MPa.} The occurrence of AFM order at ambient pressure (0.1\,MPa) is evidenced by the observation of a sharp kink in the magnetic susceptibility $\chi(T, p = \mathrm{const.})$ below which an easy-plane anisotropy evolves; see Fig.\,\ref{fig:S5}\,A where the quantity $d(\chi \cdot T)/dT$ is shown. The data are consistent with previous observations reported in Refs.\,\cite{Peters2023, Wolf2023}. As discussed in more detail in Ref.\,\cite{Wolf2023}, the quantity $d(\chi \cdot T)/dT$ was found to be proportional to the electronic specific heat. Based on this proportionality, we identify the midpoint of the jump-like anomaly (dashed line in Fig.\,\ref{fig:S5}\,A) with the temperature of AFM ordering, resulting in $T_\mathrm{N} = 47$\,K. \newline
In Fig.\,\ref{fig:S5}\,B we show data for the in-plane coefficient of thermal expansion, $\alpha_{a}(T)$ = $1/L_a \cdot \left( d L_a/d T \right )_p$, on the same temperature scale. The data show a gradual but strong increase in $\alpha_{a}(T)$ upon cooling from 100\,K down to 47\,K, where it adopts a pronounced sharp maximum, followed by an abrupt decrease. A comparison with the magnetic data in panel (A) reveals that it is the position of the maximum in $\alpha_{a}(T)$ (dashed vertical line) that corresponds to $T_\mathrm{N}$. For temperatures $T < T_\mathrm{N}$, $\alpha_{a}(T)$ is small (compared to the behavior at $T > T_\mathrm{N}$), and shows a weak negative anomaly slightly below 40\,K. As argued in the main text, we assign the strong increase in $\alpha_{a}(T)$ with decreasing temperature above $T_\mathrm{N}$ to contributions from valence fluctuations that grow upon cooling. However, this effect abruptly disappears by cooling through $T_\mathrm{N}$. \newline
In panel (C) we show for comparison resistivity data $\rho(T, p = 0.1\,\mathrm{MPa})$, plotted as $-\rho^{-1} d\rho/dT$. In this representation, the data show essentially the same behavior as $\alpha_{a}(T)$, with a pronounced peak at $T_\mathrm{N} = 47$\,K and a strong reduction on the low-temperature flank. \newline 
 \newline \clearpage
\noindent \textbf{Valence-crossover behavior at $p$ = 81\,MPa.}In Fig.\,\ref{fig:S6}\,A we show the quantity $d(\chi \cdot T)/dT$ at nominal $p= 81$\,MPa. In contrast to the sharp kink in the data at $p = 0.1$\,MPa, and the anisotropic magnetic response below, the data at 81\,MPa reveal a more rounded peak around 49\,K. In particular, measurements in $H$ = 10\,kOe for fields along and perpendicular to the $c$ axis reveal practically identical results. As discussed in Ref.\,\cite{Wolf2023}, this behavior is characteristic of a valence crossover, manifesting itself in a Lorentzian-shaped maximum in $d(\chi \cdot T)/dT$ at the valence-crossover temperature $T'_{\mathrm{V}}$, resulting in $T'_{\mathrm{V}}$ = 49\,K. The weak shoulder on the low-temperature flank of the maximum is assigned to a small volume fraction of the crystal with a somewhat reduced $T'_{\mathrm{V}}$. \newline
In panel (B), we show corresponding data at nominal $p$ = 81\,MPa for the in-plane coefficient of thermal expansion $\alpha_{a}(T)$ = $1/L_a \cdot \left( d L_a/d T \right )_p$ on the same temperature scale. The data show essentially the same behavior as $d(\chi \cdot T)/dT$, i.e., a main, slightly rounded maximum at $T'_{\mathrm{V}}$ = 49\,K and a second, less pronounced peak on the low-temperature flank.  \newline 
In panel (C) we show resistivity data $\rho(T)$ at nominal $p$ = 81\,MPa, plotted as $-\rho^{-1} d\rho/dT$. The data reveal a single peak at $T'_{\mathrm{V}}$ = 49\,K but lack indications for a second anomaly on the low-temperature flank. This observation is consistent with the assignment of the second anomaly in panels (A) and (B) to reflect valence-crossover behavior with reduced $T'_{\mathrm{V}}$ originating from a small volume fraction that is below the percolation threshold. 
\clearpage
\noindent \textbf{Competition of AFM order and VF for 0.1\,MPa $< p <$ 30\,MPa.} In  Fig.\,\ref{fig:S7}\,A, we show data of the coefficient of thermal expansion $\alpha_a (T, p = \mathrm{const.})$ as a function of temperature at varying constant pressure values. These data, derived from the $\Delta L_a/L_a$ data shown in Fig.\,3 in the main text through numerical differentiation, reflect the bifurcation into two different ground states depending on the applied pressure: an antiferromagnetically ordered state is obtained for $p \leq$ 14\,MPa. This state is characterized by a small $\alpha_a$ close to the zero line in Fig.\,\ref{fig:S7}\,A, see the comment below on the negative contribution. For higher pressures of $p \geq$ 15\,MPa, the data reveal a large positive contribution to $\alpha_a$, trending towards a universal behavior below about $30$\,K, indicative of a state governed by strong valence fluctuations.\newline
At temperatures above the bifurcation, all data sets show a more or less pronounced positive upturn upon cooling below $60$\,K, an effect that increases with increasing pressure. This upturn is assigned to valence fluctuations, which are already present to some extent at $p$ = 0.1\,MPa, but significantly gain in strength under pressure, as reflected in the large pressure coefficient $dT'_{\mathrm{V}}/dp$ = (90 - 100) K/\,GPa. At low pressures $p \leq$ 14\,MPa, however, the trend toward magnetic order prevails and the effect of valence fluctuations becomes strongly suppressed by the onset of AFM order. As detailed in Fig.\,\ref{fig:S5}, the position of the maximum in $\alpha_a$ can be assigned to the ordering temperature $T_\mathrm{N}(p)$, see the arrows in Fig.\,\ref{fig:S7}\,B. A remnant of this maximum is still visible in the data at higher pressures of 15\,MPa and above, where it gives rise to a shoulder-like anomaly. We interpret this effect as an incipient magnetic ordering which disintegrates, however, as a result of the growing strength of valence fluctuations upon further cooling. \newline 
Figure \ref{fig:S8} shows results of the magnetic susceptibility plotted as $d(\chi \cdot T)/dT$ for magnetic fields of $H$ = 10\,kOe applied perpendicular and parallel to the $c$ axis. The data taken at a $p$ = 14\,MPa (Fig.\,\ref{fig:S8}\,A) reveal the clear signatures of AFM ordering, that is, a jump-like anomaly in $d(\chi \cdot T)/dT$ at $\sim$ 46\,K, below which a magnetic anisotropy evolves. At this pressure and for $T <$ 41\,K, the sample is in the regime of phase coexistence (Fig.\,\ref{fig:S9}), where both phases contribute to the magnetic signal according to their relative volume fraction (which changes with temperature). In contrast, an almost isotropic behavior in $d(\chi \cdot T)/dT$ is revealed for the data taken at $p$ = 29\,MPa (Fig.\,\ref{fig:S8}\,B), reflecting valence-crossover behavior consistent with the phase diagram in Fig.\,5 and Fig.\,\ref{fig:S9}. We assign the small anisotropy around 40\,K, which disappears below 20\,K, to the action of magnetic correlations, which eventually disintegrate due to the effect of strong VF. 
\newpage

\clearpage

\subsection*{E: Temperature-pressure phase diagram of \EuPdSiGex\, including anomalies of minor volume fraction with slightly different Ge concentration}
\addcontentsline{toc}{section}{Supplementary Note 5: $T$-$p$ phase diagram of \EuPdSiGex\, including ...}

\begin{figure} [b!] 
	\centering
	\includegraphics[width=0.75\textwidth]{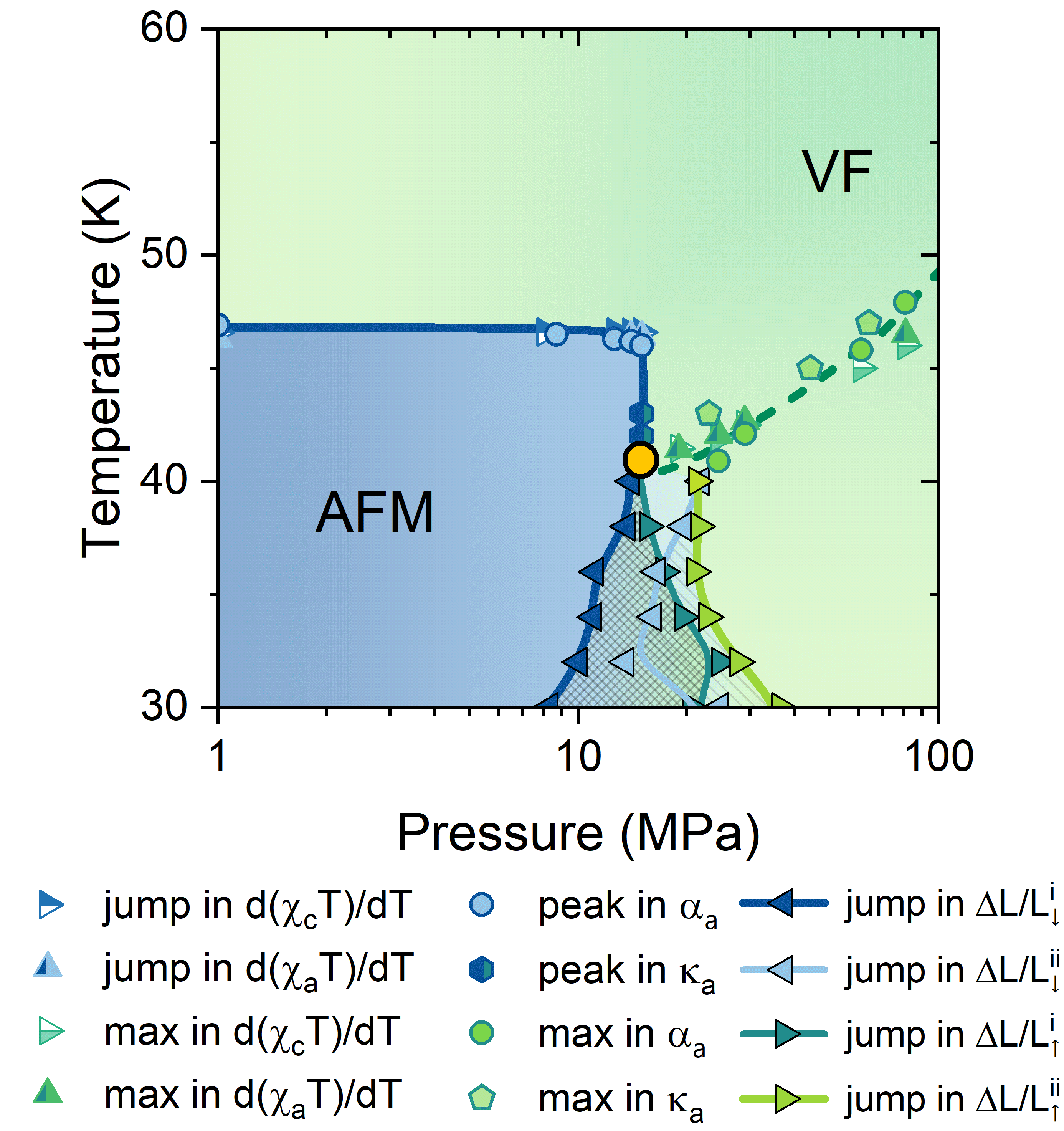} 
	\caption{\textbf{Phase diagram of \EuPdSiGex\, including signatures from majority and minority phases.} The figure replicates (in a semi-logarithmic representation) the data for $T_{\mathrm{N}}$, $T'_{\mathrm{V}}$ and the TCP (yellow-filled circle) shown in Fig.\,\ref{fig:Fig4} of the main text. In addition, for $T < T_\mathrm{TCP}$ it shows the position of the large discontinuities in $\Delta L_a/L_a$ on increasing (dark green right triangles) and decreasing (dark blue left triangles) pressure accompanying the line of first-order transitions for the major volume fraction and the position of the smaller discontinuities (light blue and light green triangles) of the minor volume fraction. The cross-hatched (left-hatched) area marks the regime of phase coexistence accompanying the line of first-order transitions for the major (minor) volume fraction. } 
        \label{fig:S9} 
\end{figure}

\begin{figure} [b!] 
	\centering
	\includegraphics[width=1\textwidth]{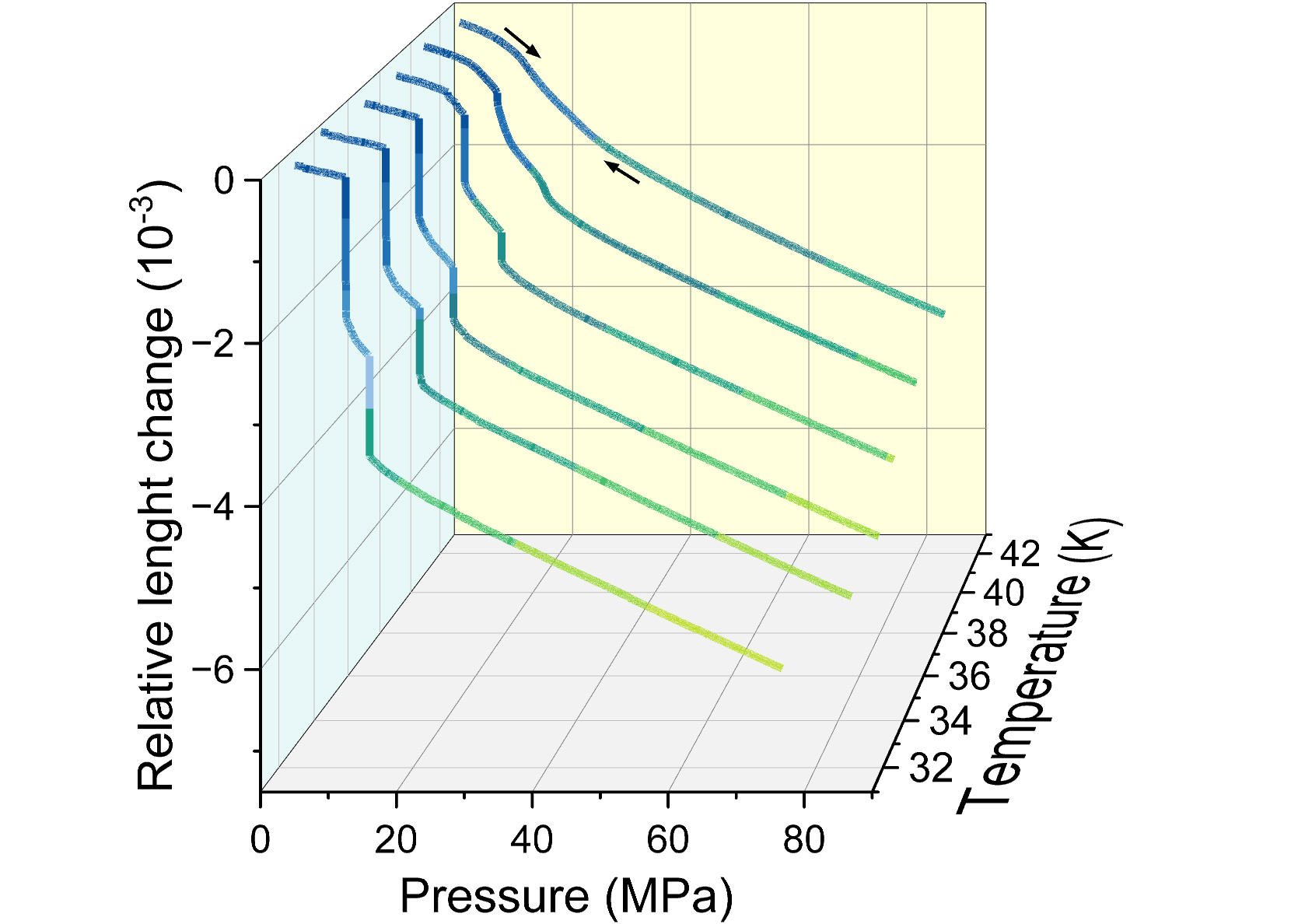} 
	\caption{\textbf{Evidence for a pressure-induced first-order transition in \EuPdSiGex\,.}  Relative length change along the $a$ axis, $\Delta L_a(p,T = {\mathrm{const.}})/L_a$ for sample \#2a, as a function of pressure at varying constant temperatures of $T$ = 42\,K, 40\,K, 38\,K, 36\,K, 34\,K, and 32\,K. Whereas the data at 42\,K are fully reversible upon increasing and decreasing pressure (arrows), hysteresis effects (not shown) are observed for the data at $T \leq$ 40\,K, see, e.g., Fig.\,\ref{fig:S11}\,A for the run at 34\,K.}
	\label{fig:S10} 
\end{figure}

In Fig.\,\ref{fig:S9} we show a replication of the $T$\,-\,$p$ phase diagram of \EuPdSiGex\, (Fig.\,5), including the TCP (yellow-filled circle) at ($p_\mathrm{TCP}, T_\mathrm{TCP}$) = $(14\pm1\,\mathrm{MPa},  41\pm1\,\mathrm{K})$. The diagram also includes the position of the large jump-like anomalies revealed in isothermal measurements of $\Delta L_a/L_a$ vs $p$ on increasing and decreasing pressure, reflecting the range of phase coexistence (cross-hatching area) accompanying the line of first-order transitions. The hysteresis closes at the TCP. The figure also includes the positions of the smaller jumps revealed in $\Delta L_a/L_a$ vs $p$ measurements on increasing and decreasing pressure, which are assigned to a minor volume fraction with slightly different Ge concentration. The corresponding hysteresis (left-hatched area) closes at a somewhat higher critical pressure of 23\,MPa.

\clearpage

\subsection*{F: Evidence for a pressure-induced first-order transition in \EuPdSiGex}
\addcontentsline{toc}{section}{Supplementary Note 6: Evidence for a pressure-induced first-order transition ...}

\begin{figure} 
	\centering
	\includegraphics[width=1\textwidth]{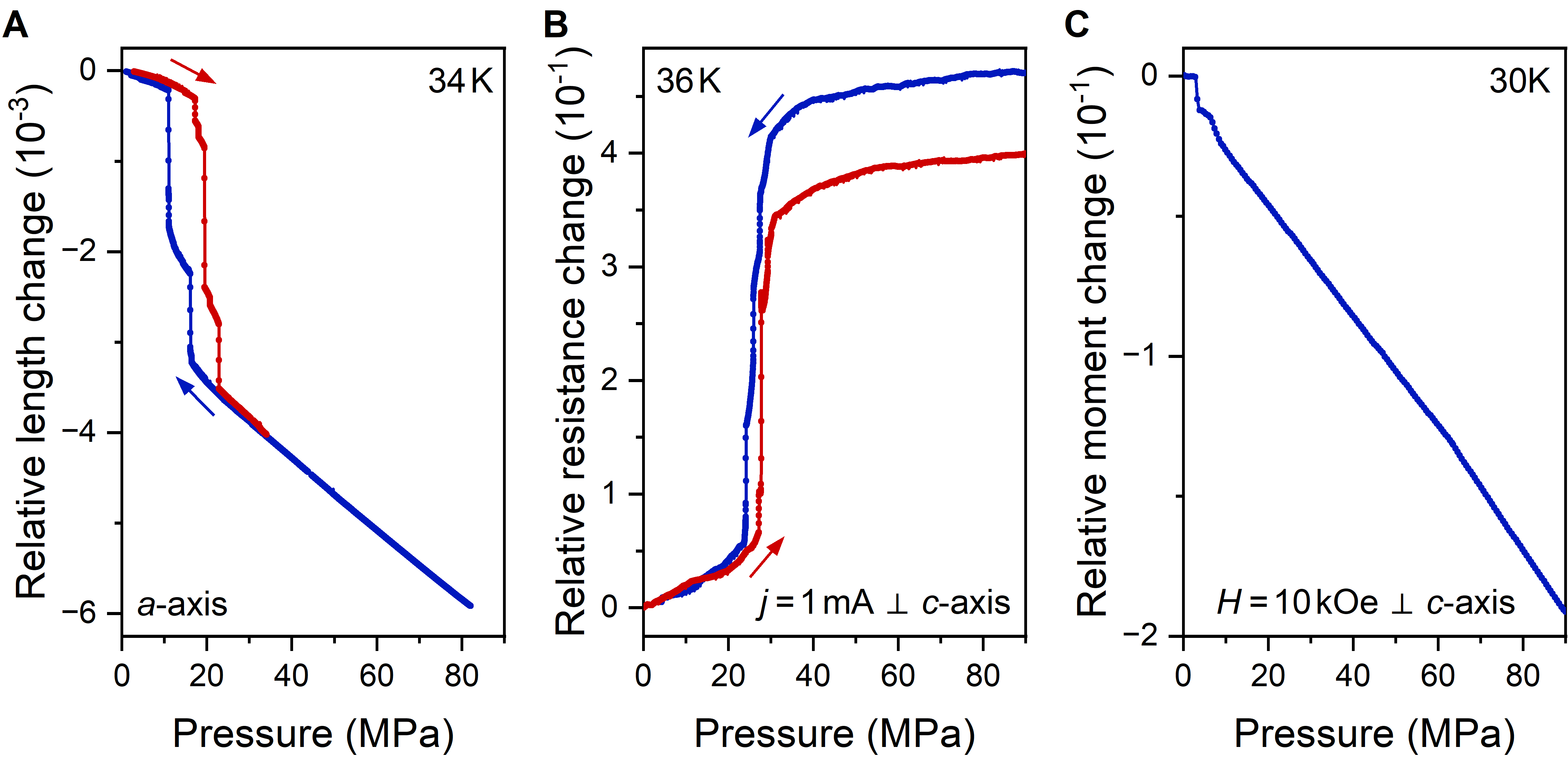} 

	\caption{\textbf{Evidence for a pressure-induced first-order transition in \EuPdSiGex\,.} 
		\textbf{(A)} Relative length change along the $a$ axis, $\Delta L_a(p,T = {\mathrm{const.}})/L_a$ for sample \#2a, as a function of pressure at $T$ = 34\,K, measured with decreasing (blue) and increasing (red) pressure. \textbf{(B)} Relative change in the resistivity, $\Delta \rho(p,T = {\mathrm{const.}})/\rho$ of sample \#2a\_1, as a function of pressure at $T$ = 36\,K, measured with decreasing (blue) and increasing (red) pressure. A current of $j=1$\,mA was applied perpendicular to the $c$ axis. \textbf{(C)} Relative change of the magnetic moment, $\Delta m(p,T = {\mathrm{const.}})/m$ for samples, as a function of pressure at $T$ = 30\,K, measured with decreasing pressure. A magnetic field of $H$ = 10\,kOe was applied perpendicular to the $c$ axis.}
	\label{fig:S11} 
\end{figure}

\textbf{Indications from pressure-dependent measurements.} Given the steepness of the phase boundary between the AFM order and the regime of strong VF in the $T - p$ plane, measurements as a function of pressure are the method of choice to probe the character of the phase transition. In Fig.\,\ref{fig:S11}\,A, we show data of the relative length change, $\Delta L_a(p,T = {\mathrm{const.}})/L_a$, along the $a$ axis as a function of pressure at $T = 34$\,K, taken with increasing (red) and decreasing (blue) the pressure. The data reveal a large jump of $\sim 0.3\%$, accompanied by hysteresis in the pressure range $11-22$\,MPa, highlighting the first-order character of the transition. In addition, changes in the slope of $\Delta L_a(p)/L_a$ for pressures below and above the critical pressure indicate a different elastic behavior: the strong increase in the slope above the critical pressure reflects a distinctly softer lattice in the VF regime as compared to the AFM state at lower pressure (cf.\,Fig.\,5 in the main text).\newline  
The first-order character of the pressure-induced phase transition is also reflected in the measurements of the resistivity and the magnetic moment. Fig.\,\ref{fig:S11}\,B shows changes of the resistivity, $\Delta \rho(p,T = {\mathrm{const.}})/\rho$, as a function of pressure at $T = 36$\,K, taken at a current of 1\,mA applied within the tetragonal plane with increasing (red) and decreasing (blue) the pressure. The offset on the high-pressure side of the transition is assigned to the occurrence of micro-cracks and/or strain fields induced by the strong lattice effects accompanying the first-order transition. \newline
Fig.\,\ref{fig:S11}\,C shows changes of the magnetic moment, $\Delta m_a(p,T = {\mathrm{const.}})/m_a$, as a function of pressure at $T = 30$\,K, measured with decreasing pressure and a magnetic field of $H$ = 10\,kOe applied within the tetragonal plane. Upon crossing the critical pressure, the magnetic moment exhibits a discontinuous change. For pressures below the critical pressure, $\Delta m_a/m_a$ is almost pressure independent on the scale of Fig.\,\ref{fig:S11}\,C. In contrast, for pressures above the critical pressure, $\Delta m/m$ decreases strongly with pressure in an almost linear fashion. Due to technical constraints, pressure sweeps could only be performed with decreasing pressure.    
 \newline
\clearpage
\newpage

\subsection*{G: Effect of pressure on resistivity}
\addcontentsline{toc}{section}{Supplementary Note 7: Effect of pressure on resistivity}

\begin{figure} [b!]
	\centering
	\includegraphics[width=0.65\textwidth]{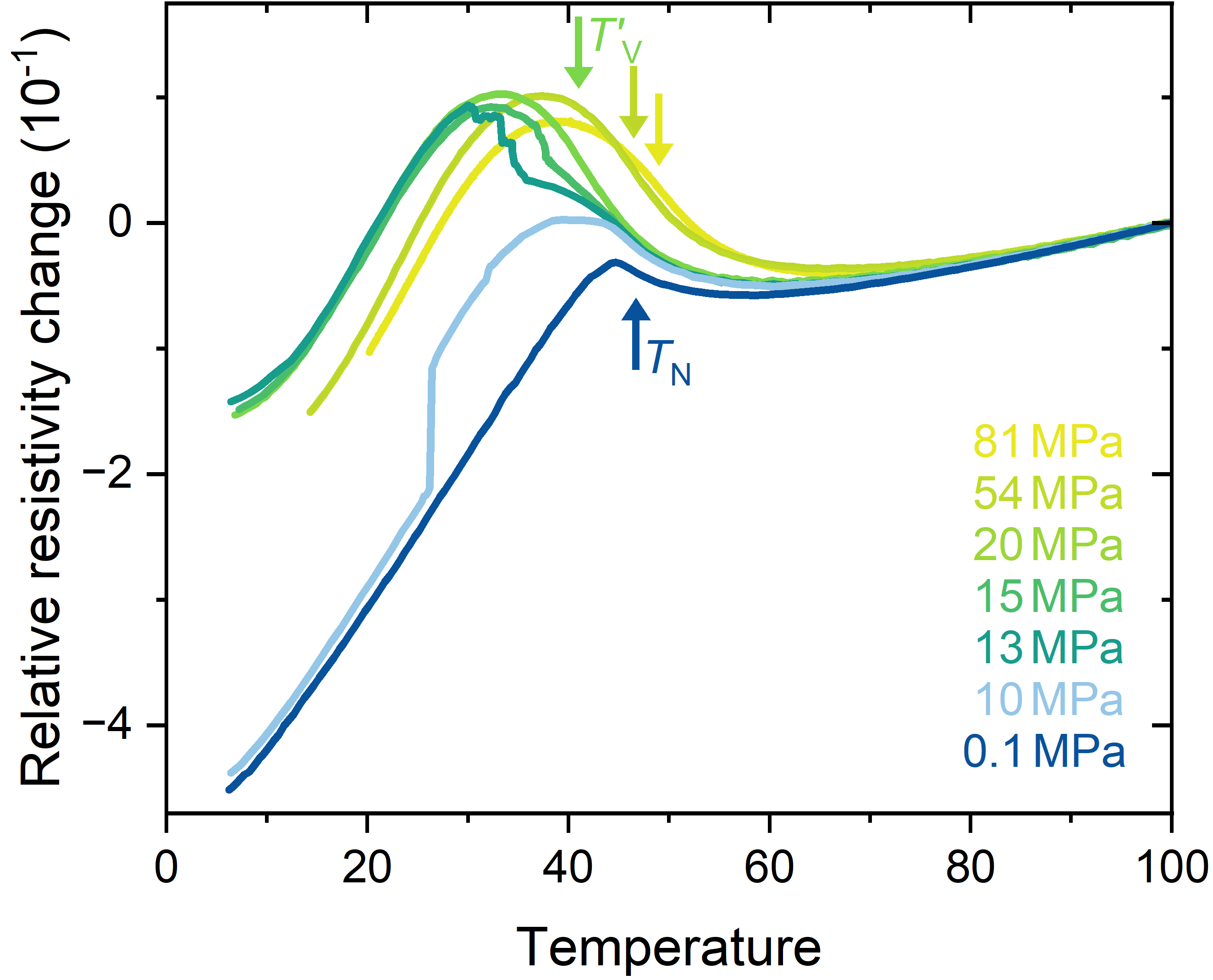} 

	\caption{\textbf{Resistivity of EuPd$_{\mathbf{2}}$(Si$_{\mathbf{0.89}}$Ge$_{\mathbf{0.11}}$)$_\mathbf{2}$ under various pressure} Relative change of the in-plane resistivity $\Delta \rho(T, p=\mathrm{const.})/\rho$ with $\Delta \rho(T, p)=\rho(T,p)-\rho(100\,\mathrm{K},p)$ as a function of temperature for sample \#2a\_2. Data were taken for varying pressure values 0.1\,MPa $\leq p \leq$ 81\,MPa. Down- and up-arrows indicate the position of the valence-crossover temperature $T'_{\mathrm{V}}$ and $T_{\mathrm{N}}$, respectively.
    }
	\label{fig:S12} 
\end{figure}

Figure \ref{fig:S12} shows changes of the in-plane resistivity $\Delta \rho(T,p=\mathrm{const.})/\rho$ as a function of temperature at selected pressures 0.1\,MPa $\leq p \leq$ 81\,MPa. The data at nominal 0.1\,MPa and 81\,MPa, reflecting the transition into AFM order (0.1\,MPa) and valence-crossover behavior (81\,MPa), respectively, were shown in Fig.\,2\,A of the main text. For pressure values $p \geq$ 20\,MPa (data colored in different green tones), i.e., sufficiently above $p_{\rm TCP}$ = (14 $\pm$ 1)\,MPa, the data reveal valence-crossover behavior characterized by a gradual increase upon cooling followed by a broad maximum. The valence-crossover temperature $T'_{\mathrm{V}}$, identified with the inflection point on the high-temperature flank of this maximum (down arrows with the same color code as the data), cf. Fig.\,\ref{fig:S6}, shows an increase with increasing pressure consistent with the results from expansivity and magnetometry, see the phase diagram in Fig.\,5. At pressures 10\,MPa $\leq p \leq$ 15\,MPa, i.e., close to $p_{\rm TCP}$, the data show the effect of valence fluctuations on cooling down to about 41\,K but exhibit clear signatures of phase separation for $T \leq T_{\rm TCP}$ = (41 $\pm$ 1)\,K manifesting itself in jump-like changes in $\Delta \rho$.  

\begin{figure} [b!]
	\centering
	\includegraphics[width=0.5\textwidth]{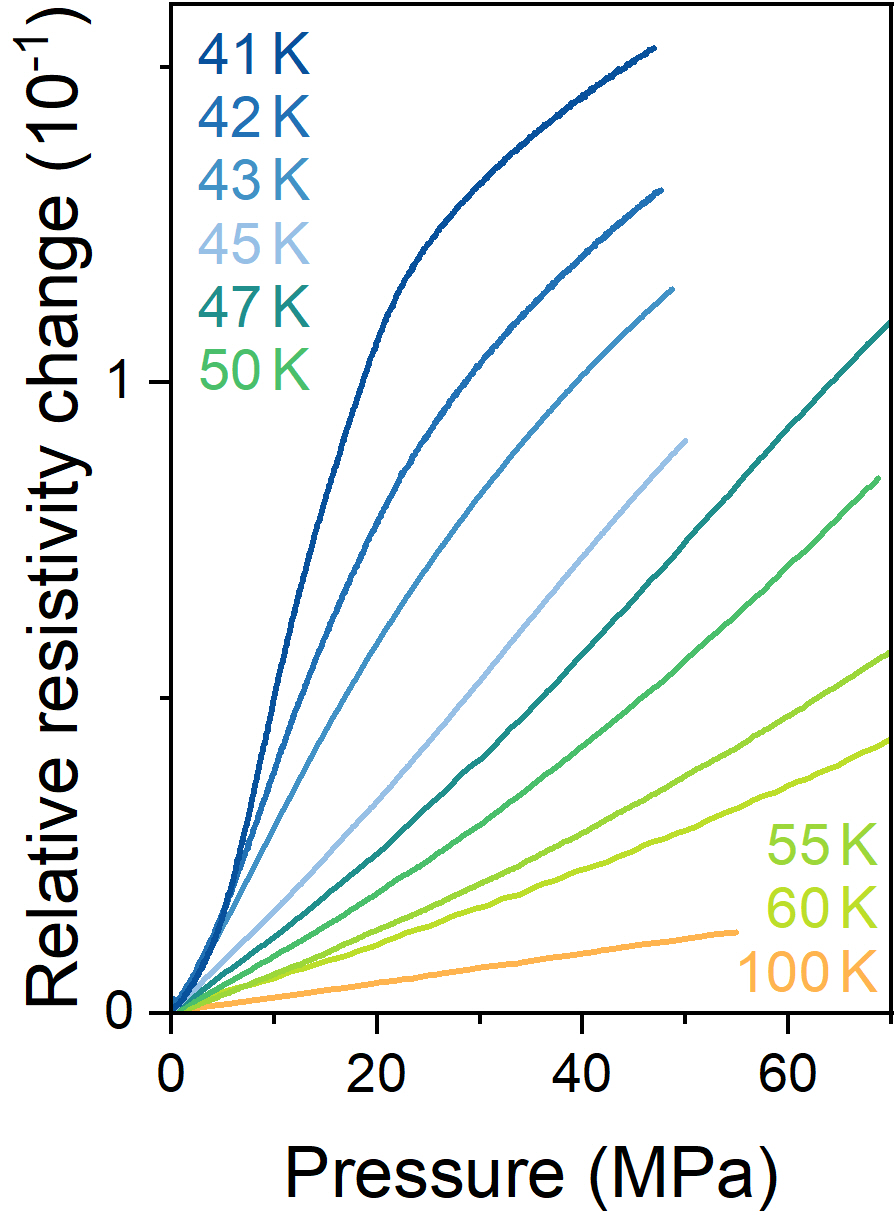} 

	\caption{\textbf{Effect of pressure on resistivity of \EuPdSiGex\,.} 
		Relative change of the in-plane resistivity $\Delta \rho(p,T=\mathrm{const.})/\rho$ as function of pressure for different constant temperatures for sample \#2a\_2. }
	\label{fig:S13} 
\end{figure}

Figure \ref{fig:S13} shows the change of the in-plane resistivity $\Delta \rho(p,T=\mathrm{const.})/\rho$ as a function of pressure at selected temperatures $T$. The data reveal strong similarities with the relative length change $\Delta L(p, T=\mathrm{const.})/L$ as a function of pressure, shown in Fig.\,2\,B, with $-\Delta \rho(p,T=\mathrm{const.})/\rho$ being almost identical to $\Delta L(p, T=\mathrm{const.})/L$. The data in Fig.\,\ref{fig:S13} were used to calculate the quantity $\kappa_{\rho}$ = $1/ \rho \cdot \left(d \rho/d p \right )_T$ shown in the main text in Fig.\,4\,B.\\

\clearpage
\newpage

\subsection*{H: Tricritical point arising in Landau theory with two interacting order parameters}
\label{Sec: H: Tricritical point arising in Landau theory with two interacting order parameters}
\addcontentsline{toc}{section}{Supplementary Note 8: Tricritical point arising in Landau theory with two ...}

Motivated by the interplay of magnetic and valence ordering, we consider a Landau theory with two interacting order parameters and demonstrate that a tricritical point (TCP) emerges in an extended parameter regime. Whenever the location of the critical endpoint (CEP) of the valence transition is sufficiently close to the magnetic transition, or, equivalently, the interaction is sufficiently strong, a TCP arises. We discuss the resulting phase diagram and determine the corresponding threshold parameters. 

The magnetic ordering is assumed to be described by an order parameter $\boldsymbol \phi$ governed by the Landau potential
\begin{equation}
    f_\phi = \frac{r}{2} \boldsymbol \phi^2 + \frac{u}{4!} (\boldsymbol \phi^2)^2 + \frac{v}{6!} (\boldsymbol \phi^2)^3 .
\end{equation}
We assume $u > 0$ and $v>0$ such that magnetic transition without an interaction with valence order is located at $r=0$. The valence ordering is represented by the Ising order parameter $\psi$. As its Ising symmetry is only emergent, the corresponding Landau potential in principle contains all powers of $\psi$. Keeping only terms up to order $\psi^4$ we get after eliminating the $\psi^3$ term by an appropriate shift $\psi \to \psi$ + const.,
\begin{equation}
    \label{eq:psi-landau-theory}
    f_\psi = a \psi + \frac{b}{2} \psi^2 + \frac{d}{4!} \psi^4.
\end{equation}
For a positive parameter $d >0$, this potential exhibits a first-order valence transition for $b < 0$ upon tuning the parameter $a$ through zero. The critical endpoint is located at $a = b = 0$.
Adding the lowest order coupling between the two order parameters, we obtain the total Landau potential,
\begin{equation}
    \label{eq:two-component-landau-theory}
    f = f_\phi + f_\psi + \beta \psi \boldsymbol \phi^2,
\end{equation}
with the coupling constant $\beta$. Due to the time-reversal symmetry, the valence order parameter $\psi$  can only couple to the square of the magnetic order parameter $\boldsymbol \phi^2$.

\begin{figure}
	\centering
	\includegraphics[width=0.95\textwidth]{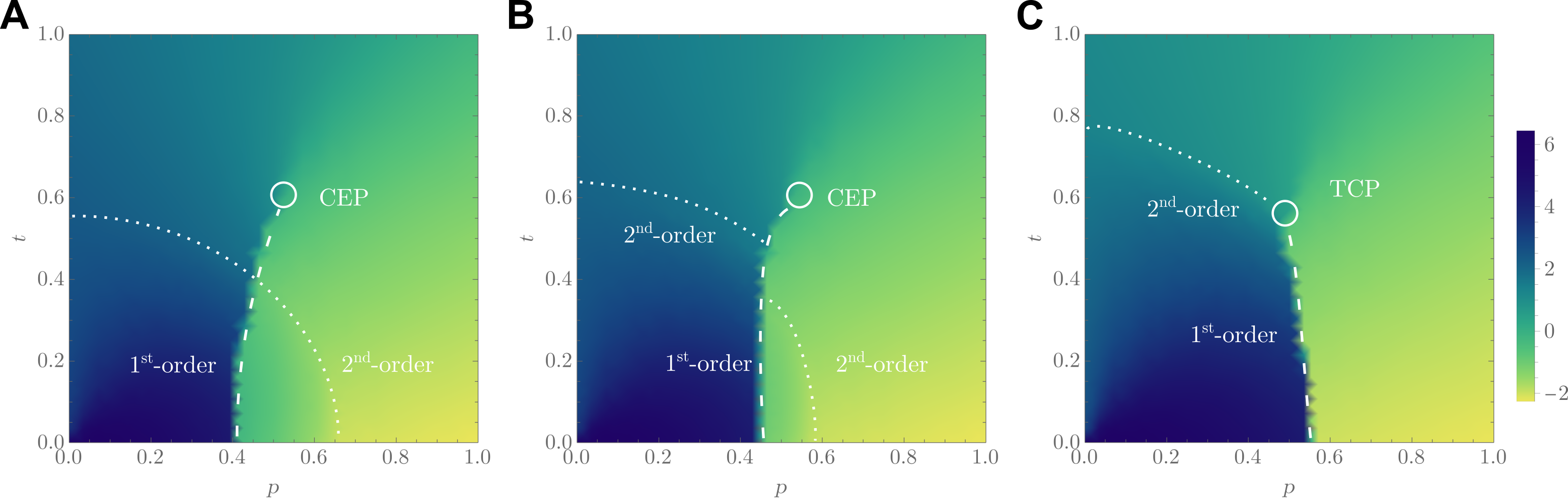} 
	\caption{
   \textbf{Phase diagram for two interacting order parameters with increasing coupling strength.} The panels show results from a numerical minimization of the Landau theory \eqref{eq:two-component-landau-theory} where the parameters are assumed to depend on the tuning parameters $p$ and $t$ on the $x$- and $y$-axis, respectively, in the form $r(t, p) = \sqrt{A t^2 + p^2} - r_c$, $a(t, p) = -(p-p_c) + B t^2$ and $b(t, p) = t-t_c$ with $d = v = 1$, $u = 0.5$, $A = 1.2$, $B = 0.3$, $t_c = 0.6$, $p_c = 0.43$, and $r_c = 0.66$.
   The color represents the first derivative of the free energy $\partial_p f$, showing clear discontinuities at the first-order transitions.
   \textbf{(A)} In the absence of an interaction, $\beta=0$, a line of second-order magnetic phase transitions (dotted line) associated with the order parameter $\boldsymbol \phi$ crosses a line of first-order valence transitions (dashed line) where the latter terminates in a critical endpoint (CEP). (\textbf{B}) A finite interaction splits the crossing point into two. (\textbf{C}) For sufficiently strong interaction a tricritical point (TCP) generically emerges where the second-order transition turns first-order. 
   }
	\label{fig:S15} 
\end{figure}

For a specific dependence of the parameters on temperature and an additional tuning parameter, the phase diagram in the absence of a coupling $\beta = 0$ is given in Fig.~\ref{fig:S15}\,A. A line of second-order phase transitions associated with magnetic order $\boldsymbol \phi$ crosses a line of first-order transitions associated with valence ordering $\psi$. In the following, we will discuss the impact of a finite coupling $\beta$ on this phase diagram. The numerical minimization shown in Fig.~\ref{fig:S15}\,B and C indicates that for a finite but weak coupling $\beta$, the crossing point between the line of first- and second-order phase transitions is split into two critical points. For larger coupling $\beta$, however, the upper critical point merges with the critical endpoint resulting in a tricritical point.

This numerical result is corroborated by analytical considerations. Minimizing the Landau potential with respect to $\psi$ we get
\begin{equation}
    \psi(\tilde a, b, d) = {\rm sign}(\tilde a) \frac{2 b - \left(\sqrt{8 b^3 + 9 \tilde a^2 d} + 3\sqrt{d}|\tilde a|\right)^{\frac23}}{
        \sqrt{d}\left(\sqrt{8 b^3 + 9 a^2 d} + 3\sqrt{d}|\tilde a|\right)^{\frac13}},
\end{equation}
where $\tilde a = a + \beta \phi^2$. Inserting this into the Landau potential, we obtain an effective Landau potential for the magnetic order parameter $\boldsymbol \phi$ only. Expanding the latter in a Taylor series in $\boldsymbol \phi$ we get
\begin{equation} \label{Eq:EffectiveLandauTheory}
    f = \frac{r'}{2} \boldsymbol \phi^2 + \frac{u'}{4!} (\boldsymbol \phi^2)^2 + \frac{v'}{6!} (\boldsymbol \phi^2)^3 + \mathcal{O}(\phi^8),
\end{equation}
with modified parameters $r'$, $u'$, and $v'$. We assume that  $v'$ remains positive,
i.e., the bare sixth-order coefficient $v$ is sufficiently large to ensure stability. For the other two parameters, we find
\begin{align}
    r' &= r + 2\beta \psi,
\\    
    u' &= u + 6\beta^2 \left(4 \psi' + (\psi')^2 (2b + d \psi^2)\right),
\end{align}
with $\psi = \psi(\tilde a, b, d)|_{\phi=0}$ and $\psi' = \partial_{\tilde a}
\psi(\tilde a, b, d)|_{\phi=0}$. The derivative $\psi'$ is negative and attains is minimum value at $a = 0$.

The interaction $\beta$ induces, in particular, a tricritical point if solutions $r' = u' = 0$ exist for some parameters. Consider the latter equation $u'=0$. It turns out that solutions exist if the coupling $\beta$ exceeds a threshold value $\beta_{\rm TCP}$. Since the correction to the quartic coupling $u'$ is largest for $a=0$ we focus on this plane in the following and evaluate  $\beta_{\rm TCP}$ for $a=0$. We obtain
\begin{align}
u'|_{a=0} = u - \frac{6\beta^2}{|b|} \times \left\{
\begin{array}{lll} 2 & {\rm for}  & b >0 \\ 
1 & {\rm for} & b < 0 .
\end{array} \right.
\end{align}
The threshold value thus scales as $\beta^2_{\rm TCP} \sim u |b|$. For a fixed bare quartic coupling, the threshold value $\beta_{\rm TCP}^2$ is proportional to $|b|$ that measures the distance to the critical endpoint at $a=0$. Consequently, Landau theory predicts a TCP provided that the coupling is sufficiently strong, $\beta^2 > \beta_{\rm TCP}^2$, or, equivalently, the CEP is sufficiently closely located to the magnetic phase transition.
In any case, an extended parameter regime arises where the coupling between valence and magnetic fluctations give rise to tricriticality. 

\clearpage
\newpage

\subsection*{I: Theory of elastic tricriticality}
\label{subsec:TheoryElasTric}
\addcontentsline{toc}{section}{Supplementary Note 9: Theory of elastic tricriticality}

The influence of a coupling between an order parameter and the elastic degrees of freedom on the critical properties of classical phase transitions has been studied intensively in the 1970ies \cite{LarkinPikin1969,Sak1974,Rudnick1974,Wegner1974,Imry1974,DeMoura1976,BergmanHalperin1976,BrunoSak1980},
see Ref.~\cite{Duenweg2000} for a review. Whereas most of these previous works focused on critical points, they also provide insights into the behavior close to tricritical points in the presence of an elastic coupling as we will demonstrate in the following. 

As a representative example, we consider a Ginzburg-Landau theory for an $N$ component order parameter field $\boldsymbol \phi$ that is coupled to elastic degrees of freedom with the Lagrangian density $\mathcal{L} = \mathcal{L}_\phi + \mathcal{L}_\varepsilon + \mathcal{L}_{\phi-\varepsilon}$. The first term comprises the Ginzburg-Landau theory
\begin{align}
\mathcal{L}_\phi = \frac{1}{2} (\partial_i {\boldsymbol \phi})^2 + \frac{u_2}{2} {\boldsymbol \phi}^2 + \frac{u_4}{4!} ({\boldsymbol \phi}^2)^2 + \frac{u_6}{6!} ({\boldsymbol \phi}^2)^3,
\end{align}
where the symbol $\partial_i$ represents the partial derivative with respect to the $i$-th spatial coordinate, and a summation over the index $i$ is implicitly assumed. In the absence of an elastic coupling, the tricritical point is attained for $u_2 = u_4 =0$ where a positive $u_6 > 0$ is required for a bounded potential. 
In the present context, this Ginzburg-Landau theory for $\boldsymbol \phi$ should be considered as the effective theory \eqref{Eq:EffectiveLandauTheory} for the magnetic order parameter after eliminating the valence fluctuations.
The second term here represents the theory of elasticity for the strain tensor $\varepsilon_{ij}$ which we assume for simplicity to be isotropic,
\begin{align}
\mathcal{L}_\varepsilon = \frac{1}{2} \Big(K - \frac{2}{3} \mu \Big) ({\rm tr}\{\varepsilon\} )^2 + \mu\, {\rm tr}\{\varepsilon^2\}, 
\end{align}
where $K$ is the bulk modulus, and $\mu$ is the shear modulus.
The third term contains the elastic coupling to the order parameter field,
\begin{align}
\mathcal{L}_{\phi-\varepsilon} = \lambda {\boldsymbol \phi}^2\,{\rm tr}\{\varepsilon\},
\end{align}
that preserves its $O(N)$ symmetry. 

Importantly, the strain tensor can be decomposed into a macroscopic part $E_{ij}$ and the microscopic phonon fluctuations described by the displacement field ${\boldsymbol u}$,
\begin{align}
\varepsilon_{ij} = E_{ij} + \frac{1}{2} (\partial_i u_j + \partial_j u_i).
\end{align}
The phonons are characterized by the longitudinal and transversal sound velocities, $c_L = \sqrt{(K + 4\mu/3)/\rho}$ and $c_T = \sqrt{\mu/\rho}$, respectively, where $\rho$ is the mass density. Due to the isotropic form of the elastic coupling, only the macroscopic volume strain, $E^{\rm vol}_{ij} = \frac{1}{3} E \delta_{ij}$, and the longitudinal phonons couple to the order parameter. 

Following Larkin and Pikin \cite{LarkinPikin1969}, an effective action is obtained by integrating out the phonon degrees of freedom
\begin{equation}
    \begin{aligned}
\mathcal{S}_{\rm eff}(E) =& \int d^d r
\Big[
\frac{1}{2} (\partial_i {\boldsymbol \phi})^2 + \frac{u_2+2 \lambda E}{2} {\boldsymbol \phi}^2 + \Big(\frac{u_4}{4!} - \frac{\lambda^2}{2\rho c_L^2} \Big) ({\boldsymbol \phi}^2)^2 + \frac{u_6}{6!} ({\boldsymbol \phi}^2)^3
\Big] \\
& + \frac{\lambda^2}{2 \rho c_L^2} \frac{1}{V} \Big(\int d^d r {\boldsymbol \phi}^2 \Big)^2 + V \frac{K}{2} E^2
\end{aligned}
\end{equation}
where $V$ is the volume. The longitudinal phonon fluctuations renormalize and effectively reduce the quartic coupling $u_4$. However, as the longitudinal phonons always carry a finite momentum, the contribution with vanishing momentum needs to be subtracted which induces a long-range interaction given by the first term in the second line. The theory describing the conditions at a given pressure, $p$, is obtained by a Legendre transformation and integrating out the volume strain $E$,
\begin{equation} \label{eq:effSp}
 \begin{aligned}
\mathcal{S}_{\rm eff}(p) =& \int d^d r
\Big[
\frac{1}{2} (\partial_i {\boldsymbol \phi})^2 + \frac{u_2 - 2\lambda p/K}{2} {\boldsymbol \phi}^2 + \Big(\frac{u_4}{4!} - \frac{\lambda^2}{2\rho c_L^2} \Big) ({\boldsymbol \phi}^2)^2 + \frac{u_6}{6!} ({\boldsymbol \phi}^2)^3
\Big] \\
& - \frac{\lambda^2}{2} \Big(\frac{1}{K} - \frac{1}{\rho c_L^2} \Big) \frac{1}{V} \Big(\int d^d r {\boldsymbol \phi}^2 \Big)^2 - \frac{V p^2}{2K}.
\end{aligned}   
\end{equation} 

This generates an effective pressure dependence for the renormalized tuning parameter $u_2$. Importantly, it also contributes to the long-range interaction that now changes its sign as $(1/K - 1/(\rho c_L^2)) > 0$.

In either case of an externally controlled strain $E$ or hydrostatic pressure $p$, the effective theory is of the form
\begin{align}
\mathcal{S}_{\rm eff}=& \int d^d r
\Big[
\frac{1}{2} (\partial_i {\boldsymbol \phi})^2 + \frac{r}{2} {\boldsymbol \phi}^2 + u ({\boldsymbol \phi}^2)^2 + \frac{u_6}{6!} ({\boldsymbol \phi}^2)^3
\Big] +  \mathrm{v} \frac{1}{V} \Big(\int d^d r {\boldsymbol \phi}^2 \Big)^2, 
\end{align}
where the sign of the long-range interaction $\mathrm{v}$ is positive or negative, respectively,
\begin{align} \label{Eq:LongRangeInt}
    \mathrm{v} = 
    \left\{ \begin{array}{lll} 
    \frac{\lambda^2}{2 \rho c_L^2} >0 & \text{for} & E = {\rm const.} \\ 
    - \frac{\lambda^2}{2} \Big(\frac{1}{K} - \frac{1}{\rho c_L^2} \Big) <0  & \text{for} & p = {\rm const.} 
    \end{array} \right.
\end{align}
The tuning parameter is either strain or pressure dependent, $r = u_2 + 2\lambda E$ or $r = u_2 - 2\lambda p/K$, respectively.
The elastic coupling $\lambda$ reduces the quartic coupling $u = u_4/4! - \lambda^2/(2 \rho c_L^2)$ thus bringing it closer to tricriticality as noted early on \cite{Aharony1973, Imry1974}. In order to determine the location of the tricritical point in the phase diagram, the theory is first considered on the mean-field level. Keeping only the homogeneous component of the order parameter field $\boldsymbol \phi_{\rm mf} = \frac{1}{V} \int d^d r \boldsymbol \phi$ one obtains the mean-field potential
\begin{align}
\mathcal{V}_{\rm eff}=& 
\frac{r}{2} {\boldsymbol \phi_{\rm mf}}^2 + (u+\mathrm{v}) ({\boldsymbol \phi_{\rm mf}}^2)^2 + \frac{u_6}{6!} ({\boldsymbol \phi_{\rm mf}}^2)^3 .
\end{align}
On the mean-field level, the tricritical point is thus located at $r = 0$ and $u = -\mathrm{v}$. However, fluctuations will in general modify  this mean-field result. 

\begin{table}[]
    \centering
    \begin{tabular}{|c|c|c|c|c|} \hline
        FP &  $u^*$ & $\mathrm{v}^*$ & $\nu$ & $\alpha$\\ \hline
        G  &  0 & 0 & $\frac{1}{2}$ & $\frac{1}{2} \epsilon$\\
        FR-G & 0 & $\frac{2\pi^2}{N}\epsilon$ & $\frac{1}{2} + \frac{\epsilon}{4}$ & $-\frac{1}{2} \epsilon$\\
        WF   & $\frac{2\pi^2}{N+8}\epsilon$ & 0 & $\frac{1}{2} + \frac{2+N}{4(N+8)} \epsilon$  & $\frac{4-N}{2(N+8)} \epsilon$\\
        FR-WF & $\frac{2\pi^2}{N+8}\epsilon$ & $\frac{2\pi^2 (N-4)}{N(N+8)}\epsilon$ & 
        $\frac{1}{2} + \frac{3}{2(N+8)} \epsilon$  & $\frac{N-4}{2(N+8)} \epsilon$ \\
        \hline
    \end{tabular}
    \caption{\textbf{Four fixed points of the renormalization group flow given in Eqs.~\eqref{SakRGu} and \eqref{SakRGv}}.
    $u^*$ and $\mathrm{v}^*$ are the fixed point values of the flowing coupling constants, $\nu$ is the correlation length exponent and $\alpha$
    is the specific heat exponent \cite{Sak1974}.
    }
    \label{tab:SakRG}
\end{table}

\begin{figure} 
	\centering
	\includegraphics[width=0.48\textwidth]{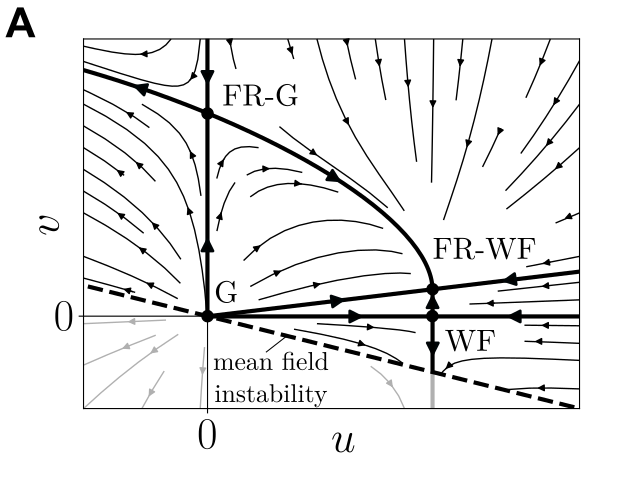} \hfill
	\includegraphics[width=0.48\textwidth]{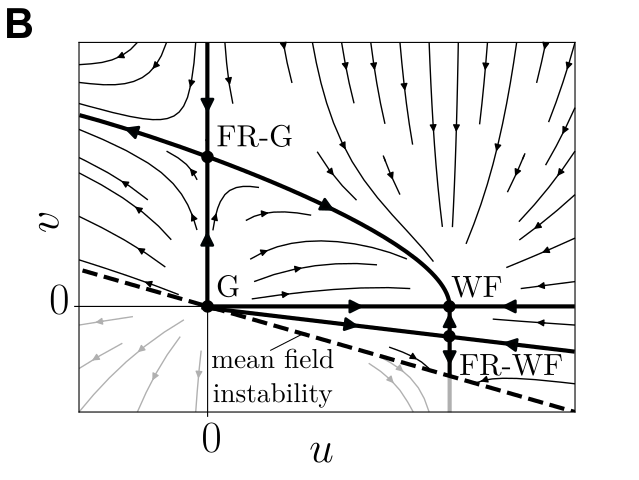} 

	\caption{\textbf{Renormalization group flow given in Eqs.~\eqref{SakRGu} and \eqref{SakRGv}.} The couplings $u$ and v are the short- and infinite-range interaction of the order parameter, where the latter is induced by the elastic coupling. The dashed line represents the mean-field instability, $u+\mathrm{v} = 0$. There exist four fixed points: the Gaussian (G) and the Wilson-Fisher (WF) fixed points and their Fisher renormalized variants (FR-G) and (FR-WF). The RG flow depends on the number of components $N$ of the order parameter and differs qualitatively for $N<4$ (panel \textbf{A}) and for $N>4$ (panel \textbf{B)}).
    }
	\label{fig:SakRG} 
\end{figure}

The influence of fluctuations was considered by Sak \cite{Sak1974} in the framework of an renormalization group (RG) treatment close to the upper critical dimension $d = 4-\epsilon$. At lowest order in $\epsilon>0$, the following RG flow for the relevant parameters $r$, $u$ and $\mathrm{v}$ are found
\begin{align}
\frac{\partial r}{\partial \ell} &= 2 r - \frac{N+2}{2 \pi^2} u r - \frac{N}{2\pi^2} \mathrm{v} r,\\
\label{SakRGu}
\frac{\partial u}{\partial \ell} &= \epsilon u - \frac{N+8}{2\pi^2} u^2, \\
\label{SakRGv}
\frac{\partial \mathrm{v}}{\partial \ell} &= \epsilon \mathrm{v} - \frac{N+2}{\pi^2} u \mathrm{v} - \frac{N}{2\pi^2} \mathrm{v}^2.
\end{align}
There are four fixed points in the $(u,\mathrm{v})$ parameter plane, see Table \ref{tab:SakRG}.
The corresponding RG flow is depicted in Fig.~\ref{fig:SakRG}. 
In the absence of an elastic coupling $\mathrm{v} \equiv 0$ one recovers the unstable Gaussian fixed point (G) and the stable Wilson-Fisher fixed point (WF). The stable WF governs the critical behavior of the second-order phase transition, whereas the unstable Gaussian fixed point governs the tricritical behavior. The fixed point value of the specific heat exponent $\alpha  = \epsilon/2$ for the Gaussian tricritical point is exact even in $d=3$ where $\alpha = 1/2$ \cite{Riedel1972}.
In the presence of an elastic coupling, $\mathrm{v} \neq 0$, however, two additional fixed points arise, (FR-G) and (FR-WF). Their specific heat exponents is consistent with Fisher renormalization (FR) \cite{Fisher1968}, $2-\alpha_{\rm FR} = \frac{2-\alpha}{1-\alpha}$, of $\alpha$ for the Gaussian and Wilson-Fisher fixed  points, respectively, see Table \ref{tab:SakRG}.
From general arguments given in the main text (see the discussion section) 
it is expected that the elastic coupling is relevant provided that the specific heat exponent is positive. This is confirmed by the RG flow. As the specific heat exponent of the Wilson-Fisher fixed point changes sign as a function of the number of components $N$, see Table \ref{tab:SakRG}, the RG flow differs for $N<4$ and $N>4$.  

For $N<4$, see Fig.~\ref{fig:SakRG}\,A, the specific heat exponent of the Gaussian and the Wilson-Fisher fixed point is positive and, thus, both are unstable with respect to the elastic coupling. There exist only a single stable critical point FR-WF. As this fixed point is located in the half-plane $\mathrm{v} > 0$, it will govern the behavior of the second-order phase transition under the condition of a fixed volume strain $E$, see \eqref{Eq:LongRangeInt}. In the same half-plane there exists the fixed point FR-G that is unstable with respect to a finite $u \neq 0$. 
Whereas for positive $u$ the flow is towards to the WF fixed point, for negative $u$ the flow is towards values with $-u \gg \epsilon$. Consequently, the fate of the RG flow is unclear within the one-loop RG treatment. Nevertheless, it is suggestive that the FR-G might capture the tricritical behavior at fixed volume strain $E$.
%
Under the condition of fixed hydrostatic pressure $p$, the long-range coupling is negative, $\mathrm{v} < 0$, and the flow is towards the mean-field instability line. In this case, the transition is always first order and a tricritical point does not exist. 

In contrast, for $N>4$ the WF fixed point is stable and it governs the second-order phase transition, see Fig.~\ref{fig:SakRG}\,B. 
As the initial value of $u$ decreases at negative $\mathrm{v}$, i.e., under the condition of a fixed pressure $p$, the separatrix is approached that connects the Gaussian with the FR-WF fixed point, which is now located in the lower half plane. It can be confirmed by a straightforward calculation that this separatrix is given by a straight line with slope $\mathrm{v}/u = - (N-4) /N$. On this separatrix the flow is towards the FR-WF and beyond the separatrix the flow is towards the mean-field instability line $\mathrm{v}/u = -1$. The FR-WF fixed point thus governs the tricritical behavior in the presence of an elastic coupling under the conditions of a fixed hydrostatic pressure $p$, see also the discussion in Ref.~\cite{Wegner1974}. Its specific heat exponent $\alpha = \frac{N-4}{2(N+8)} \epsilon$ is positive for $N>4$. As the most relevant operator $r$ depends on pressure $p$, see \eqref{eq:effSp}, the compressibility will diverge with the same exponent, $\kappa \sim |r|^{-\alpha}$, at the FR-WF tricritical point.

For the experimental setup of interest, the system is controlled with pressure $p$, and the FR-WF fixed point in Fig.~\ref{fig:SakRG}\,B is a potential candidate for the explanation of the tricritical behavior observed experimentally. 
Note that for constant pressure $p$, the long-range interaction is only finite due to the presence of a finite shear modulus $\mu$ as $\mathrm{v} = - \frac{\lambda^2}{2} \Big(\frac{1}{K} - \frac{1}{K+\frac{4 \mu}{3}} \Big) = - \frac{2\lambda^2 \mu}{3 K(K+\frac{4 \mu}{3})}$, see \eqref{Eq:LongRangeInt}. The shear modulus, that distinguishes a liquid from a solid, is thus instrumental for governing the elastic tricriticality at constant $p$. 
Despite the simplifying assumption of isotropic elasticity and the RG treatment being restricted to first order in $\epsilon$, the above discussion elucidates the concept of elastic tricriticality where the presence of an elastic coupling destablizes the Gaussian tricritical point, leads to the emergence of new tricritical fixed point and, therefore, modifies the tricritical universality class. The resulting divergence of the compressibility at elastic tricriticality is a fingerprint of the non-perturbative hybridization between critical and elastic degrees of freedom.



\newpage
\nolinenumbers


\addcontentsline{toc}{section}{Supplementary References}

\end{document}